\documentclass{aa} 

\usepackage{graphicx}
\usepackage{txfonts}
\usepackage{hyperref}
\hypersetup{
 colorlinks=true,
 urlcolor=blue,
 linkcolor=blue,
 citecolor=blue
}
\usepackage{adjustbox}
\usepackage{blindtext}
\usepackage{upgreek}
\usepackage{mathrsfs}
\usepackage{float}
\usepackage{xcolor}
\usepackage{url}
\usepackage{natbib}
\usepackage{textcomp}
\usepackage{multirow}
\usepackage{enumitem}

\newcommand\CaII{\ion{Ca}{II}}
\newcommand\CaIIK{\CaII~K}
\newcommand\CaIIH{\CaII~H}
\newcommand\CaIIHK{\CaII~H\&K}
\newcommand\CaIIIR{\CaII~8542~\AA}
\newcommand\MgII{\ion{Mg}{II}}
\newcommand\MgIIK{\MgII~k}

\newcommand\Fe{\ion{Fe}{I}}
\newcommand\FeI{\Fe~6173~\AA}
\newcommand\Halpha{{H$\alpha$}}
\newcommand\logt{{\log(\tau_{500\,\mathrm{nm}})}}

\newcommand\kms{\mbox{km s$^{-1}$}}
\newcommand\vpos{$v_{\rm{POS}}$}
\newcommand\vlos{$v_{\rm{LOS}}$}
\usepackage{ulem}
\usepackage{color}
\definecolor{deepmagenta}{rgb}{0.8, 0.0, 0.8}

\newcommand{\rf}[1]{{\textnormal{#1}}}

\begin{document} 

 \title{Transverse oscillations in 3D along Ca II K bright fibrils in the solar chromosphere}


 \author{
 Sepideh Kianfar\inst{1} 
 \and
 Jorrit Leenaarts\inst{1}
 \and
 Sara Esteban Pozuelo\inst{1,2,3}
 \and
 Jo\~{a}o M. da Silva Santos\inst{4}
 \and
 Jaime de la Cruz Rodr\'{i}guez\inst{1}
 \and
 Sanja Danilovic\inst{1}
 }

 \institute{Institute for Solar Physics, Dept. of Astronomy, Stockholm University, Albanova University Center, 10691 Stockholm, Sweden
 \and
 Instituto de Astrof\'isica de Canarias, C/V\'ia L\'actea s/n, 38205 La Laguna, Tenerife, Spain
\and
Departamento de Astrof\'{i}sica, Universidad de La Laguna, 38206 La Laguna, Tenerife, Spain
\and
National Solar Observatory, 3665 Discovery Drive, Boulder, CO 80303, USA
} 
 \offprints{Sepideh Kianfar \email{sepideh.kianfar@astro.su.se}}
 
 \date{Received Month dd, yyyy; accepted Month dd, yyyy}

 
 \abstract
 {Fibrils in the solar chromosphere carry transverse oscillations as determined from non-spectroscopic imaging data. They are estimated to carry an energy flux of several kW~m$^{-2}$, 
 which is a significant fraction of the average chromospheric radiative energy losses.} 
 {We aim to determine oscillation properties of fibrils not only in the plane-of-the-sky (horizontal) direction, but also along the line-of-sight (vertical) direction.} 
 {We obtained imaging-spectroscopy data in \FeI, \CaIIIR, and \CaIIK\ with the Swedish 1-m Solar Telescope. We created a sample of \rf{605} bright \CaIIK\ fibrils and measured their horizontal motions. Their vertical motion was determined through non-LTE inversion of the observed spectra. We determined the periods and velocity amplitudes of the fibril oscillations, as well as phase differences between vertical and horizontal oscillations in the fibrils.} 
 {The bright \CaIIK\ fibrils carry transverse waves with a mean period of \rf{$2.1\times10^2$~s}, and a horizontal velocity amplitude of \rf{1~\kms}, consistent with earlier results. The mean vertical velocity amplitude is \rf{1.1}~\kms. We find that \rf{77\% of the} fibrils carry waves in both the vertical and horizontal directions, and \rf{80\% of this subsample exhibits oscillations with similar periods in both horizontal and vertical directions.} For \rf{the latter}, we find that all phase differences between $0$ and $2\pi$ occur, with a mild but significant preference for linearly polarized waves (phase difference of $0$ or $\pi$).} 
 {The results are consistent with the scenario where transverse waves are excited by granular buffeting at the photospheric footpoints of the fibrils. Estimates of transverse wave flux based only on imaging data are too low because they ignore the contribution of the vertical velocity.} 
 \keywords{Sun: atmosphere -- Sun: chromosphere -- Sun: oscillations -- Methods: observational}

 \maketitle
%
\section{Introduction}
\label{s:intro}

The solar chromosphere is pervaded by waves and oscillations that are excited by the convection and p-modes in the photosphere
\citep[e.g.,][and references therein]{2015SSRv..190..103J}. 
These waves are considered prime candidates for transporting the energy needed to sustain chromospheric radiative losses from the photosphere into the chromosphere. 

In areas where the gas pressure is stronger than the magnetic pressure (plasma $\beta>1$) acoustic-gravity waves dominate. They are well-understood, both observationally and theoretically
\citep[e.g.,][]{1997ApJ...481..500C,2004A&A...414.1121W},
but the extent of their contribution to the required energy input remains under debate
\citep{2019ARA&A..57..189C}.
In areas where the magnetic pressure is stronger ($\beta<1$), the situation is more complex. In a homogeneous plasma there are now three waves, the slow and fast magneto-acoustic wave, and the Alfv{\'e}n wave. The chromosphere is obviously not homogeneous, and a large literature exists about waves in homogeneous magnetic cylinders (referred to as flux tubes in the literature) with the field aligned to the axis of the cylinder, taken as an approximation of magnetic field bundles 
\citep[e.g.,][]{1983SoPh...88..179E,Zaqar2009,2009A&A...503..213G}.
In this geometry there is a complex set of wave modes, including torsional waves, oscillations of the tube diameter (sausage modes), and a swaying of the whole tube (kink modes).
\rf{The kink mode is the most frequently identified oscillation in observations,}
as it manifests itself as clearly swaying structures in imagery, especially of spicules that protrude above the limb of the Sun
\citep[e.g.,][]{2007Sci...318.1574D,2008ASPC..397...27S}.
%


Propagating transverse oscillations of fibrils and mottles, elongated chromospheric structures that emanate from photospheric magnetic elements have been observed using imaging data in chromospheric spectral lines, e.g.,
\citep{2011ApJ...739...92P,2012ApJ...750...51K,2014ApJ...784...29M,jafarzadeh17_2}.
These oscillations have typically been interpreted as kink waves. Observed velocity amplitudes are on the order of 5~\kms, periods in the range of 16\,--\,600~s, and phase speeds of 50\,--\,300~\kms.
\citet{2021ApJ...921...30S} studied high frequency waves in spicule-like event using imaging spectroscopy on \Halpha. In addition to plane-of-the-sky (POS) velocities, they used Doppler shift information to estimate line of sight (LOS) velocity of the spicules. They found periods in the range of 
40\,--\,200~s, and velocity amplitudes around 5--20~\kms.

However, when radial and longitudinal inhomogeneities are introduced in magnetic concentrations, the separation between the different wave types becomes more difficult
\citep{2021JGRA..12629097S}. 
Radiation-MHD models of the chromosphere show proxies of different wave types often appear in the same location and become indistinguishable 
\citep{2022arXiv220803744D}. 
\citet{jorrit15} 
investigated the relation between oscillations as seen in \Halpha\ imagery and transverse oscillations propagating along magnetic field lines in a radiation-MHD simulation of a network. They warned that observed fibrils might not always trace out single field lines, so that interpretation of observations in terms of kink waves along flux tubes must be done with caution. However, radiative transfer computations by
\citet{2019A&A...631A..33B} 
based on a radiation-MHD simulation of an active region, indicate that strong fibrils seen in \Halpha, \MgIIK, and \CaIIK\ trace field lines to a much larger degree than in the network simulations.

%

Several observations of transverse oscillations in off-limb spicules using both POS swaying and LOS motions through measuring Doppler shifts of line profiles exist
\citep[e.g.,][]{1982SvAL....8..341G,2012ApJ...752L..12D,2018ApJ...856...44A}
, but only one where on-limb fibrils are considered \citep{2021ApJ...921...30S}. In the latter study, Doppler shifts in H$\alpha$ are used to trace the LOS variations in velocity, which significantly limits the accuracy.
%

In this paper we exploit imaging spectroscopy data obtained at the Swedish 1-m Solar Telescope in multiple spectral lines combined with non-LTE inversion to trace the LOS components \rf{as well as the POS components }of transverse oscillations. \rf{Therefore, we provide a thorough study of the oscillations in 3D which ensures that the LOS velocity is well resolved}.
We focus on oscillations in fibrils that appear bright in \CaIIHK\ data
\citep{jafarzadeh17_2,Kianfar20}.
The physical properties of these bright fibrils have been studied by \citet{Kianfar20}. We report here on both POS and LOS velocity and displacement amplitudes, wave periods, and correlations between the POS and LOS properties.

\section{Observations}
\label{s:obs}
\subsection{Data preparation}
\label{sb:data}
We analyse observations of a fibrillar area centered at $(X,Y) = [70\farcs4, 153\farcs2]$, i.e., $\mu=0.98$, taken at the Swedish 1-m Solar Telescope 
\citep[SST;][]{scharmer03}. 
The observed field of view (FoV) was located at the edge of a plage region in the AR12716. The observations were acquired on 2018-07-22, from 08:23:58 to 08:53:25~UT using the CRisp Imaging Spectro-Polarimeter \citep[CRISP;][]{crisp} and CHROMospheric Imaging Spectrometer \citep[CHROMIS;][]{CHROMIS} instruments. Seeing conditions were excellent throughout the whole observing time.

The \FeI\ line was sampled by CRISP in 13 equidistant wavelength positions in a range of $\pm 180$~m\AA\ around the line centre. CRISP also observed the \CaIIIR\ line. It was sampled at 21 equidistant wavelength positions between $\pm 550$~m\AA\, as well as two extra points at $\pm 880$~\AA\ from the line core. Both lines were observed with full Stokes polarimetry and with a total cadence of 21~s. 

CHROMIS observed the \CaIIK\ line in 27 equidistant wavelength positions in the range $\pm 1.5$~\AA\ around the line core. In addition, it observed an additional wavelength point at 4000~\AA\ (continuum). The CHROMIS observations have a cadence of 8~s.

The observed data were reconstructed by applying the CRISPRED reduction pipeline \citep{crispred} to the CRISP data and SSTRED \citep{sstred} to the CHROMIS observations. The image restoration used in the reconstruction pipelines is the Multi-Object Multi-Frame Blind Deconvolution \citep[MOMFBD;][]{momfbd} method. After reconstruction, the CRISP and CHROMIS images were co-aligned with an accuracy of about 0.1~pixel. The CRISP data were resampled from a pixel size of $0\farcs059$ to the CHROMIS pixel size of 0$\farcs$038. They were also resampled in time to the CHROMIS cadence using nearest-neighbour interpolation. \rf{The data} were calibrated to the absolute intensity by comparing the average intensity profile in a quiet region in the FoV to a solar atlas profile 
\citep{atlas}.

\subsection{Overview of the FoV} \label{sb:fov}
 \begin{figure*}
 \centering
 \includegraphics[width=\linewidth]{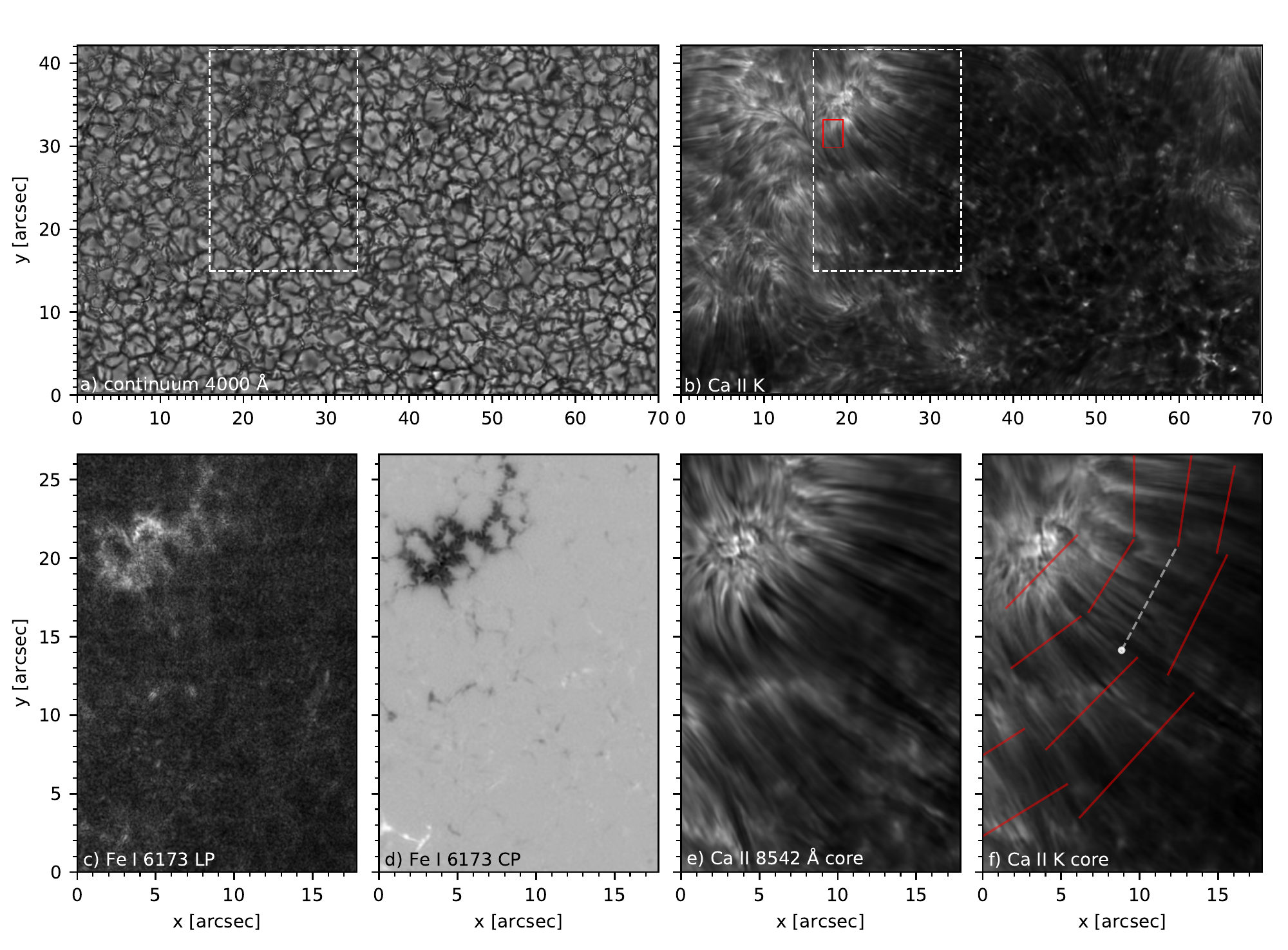}
 \caption{Overview of the observations taken on 2018-07-22 at 08:31:49 UT. \textit{Top panels} show intensity images of the continuum at 4000 \AA\ (\textit{a}) and the wavelength-integrated \CaIIK{} line intensity (\textit{b}). The white dashed box marks the region that we analyse in detail. The red box marks the region of interest (RoI) analysed in Fig~\ref{fig:roi}. A zoom-in of the area enclosed by the white dashed box is shown in the \textit{bottom panels}. \textit{c}: \FeI\ wavelength-averaged linear polarisation; \textit{d}: \FeI\ wavelength-averaged circular polarisation; \textit{e}: \CaIIIR\ line core intensity; \textit{f}: \CaIIK\ line core intensity. The red solid and white dashed lines in panel \textit{f} show the cuts across the \CaIIK\ bright fibrils that we used (see Sec.~\ref{sb:track}). An analysis of the cut along the white dashed line is presented in Figs~\ref{fig:cut_inv} and \ref{fig:cut_cat}. The white dot marks the zero point on the $y$-axis in Fig.~\ref{fig:cut_inv}. An animated version of this figure showing the entire time sequence is available online.}
 \label{fig:maps}
 \end{figure*}

Figure~\ref{fig:maps} and the accompanying animation show an overview of the observations. The upper-left corner of the FoV contains a plage region with a concentration of unipolar small-scale magnetic features as seen in the linear and circular polarization maps (Fig.~\ref{fig:maps}c and \ref{fig:maps}d). The area surrounding the plage region is covered with a forest of elongated bright fibrils in the chromosphere (Fig.~\ref{fig:maps}b). The \CaIIIR\ line core image in Fig.~\ref{fig:maps}e shows a fibrillar pattern similar to the one in \CaIIK\ \citep[cf. Fig.~\ref{fig:maps}f,][]{Kianfar20}. 
The right half of the CHROMIS FoV is quiet and therefore devoid of bright fibrils as shown in the \CaIIK\ continuum image in Fig.~\ref{fig:maps}a. We used this part of the FoV for the absolute intensity calibration of our observations (see Sect.~\ref{sb:data}).

\section{Analysis methods}
\label{s:analysis}

\subsection{Measuring POS oscillations}
\label{sb:track}

 \begin{figure}
 \centering
 \includegraphics[width=0.985\linewidth]{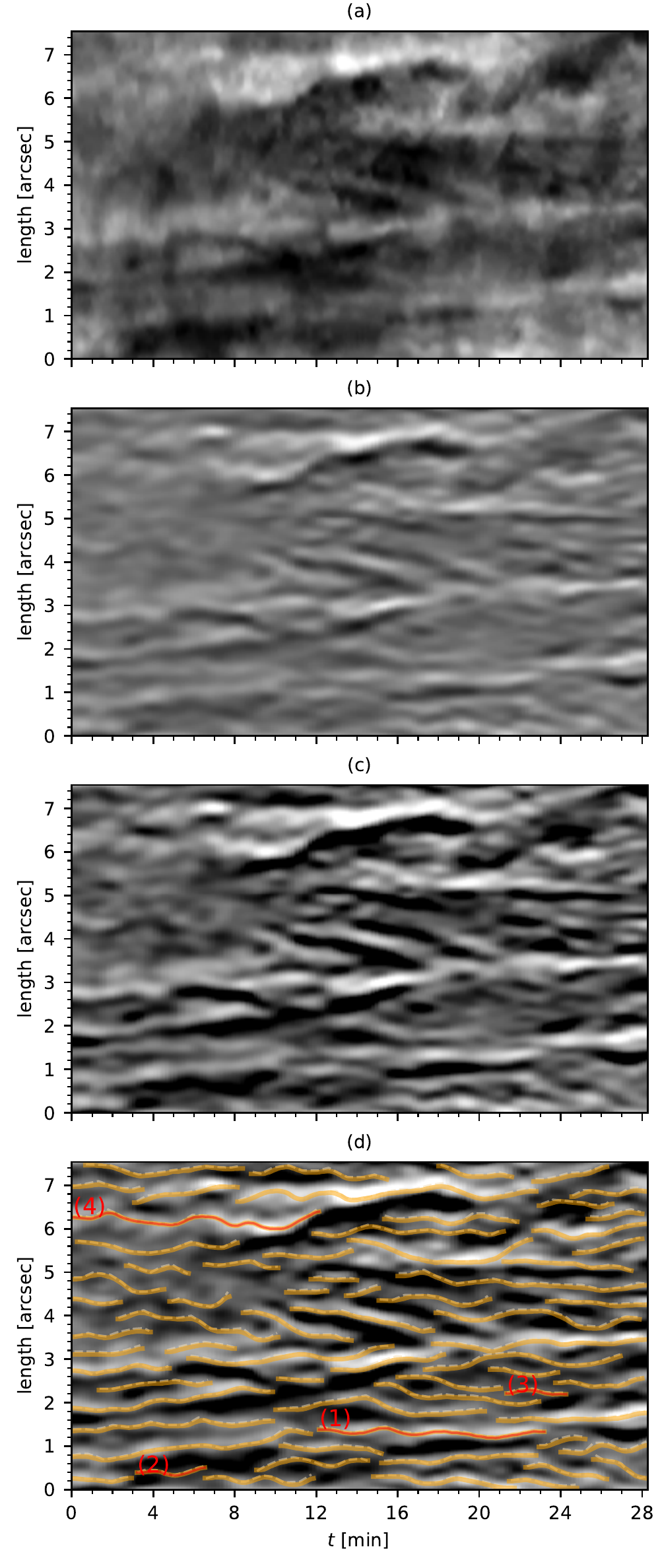}
 \caption{Time evolution of the cut across the bright fibrils shown with a white dashed line Fig.~\ref{fig:maps}-f. The \textit{panel (a)} displays the intensity at the nominal line centre of \CaIIK{}. The \textit{panel (b)} shows the enhanced intensity of \textit{panel (a)} by applying the enhancing procedure described in Sect.~\ref{sb:track}. The \textit{panel (c)} the gamma-corrected image of \textit{panel b)} that fully brings out individual fibril oscillations. The POS oscillation trajectories are marked on the high contrast space-time image in \textit{panel (d)} where the preliminary paths are plotted with dashed lines and the smoothed oscillations are plotted with solid curves. The oscillation properties of the curves marked with numbers are shown in Fig.~\ref{fig:cut_cat}. The zero point of the cut length ($y$-axis) is marked with a white dot in Fig.~\ref{fig:maps}f.}
 \label{fig:cut_int}
 \end{figure}

In order to analyse the transverse (POS) oscillations of bright fibrils visible in \CaIIK, we selected a region in our data that was observed by both CRISP and CHROMIS and contained fibrils (indicated by the white-dashed box in Fig.~\ref{fig:maps}b). Then we defined 12 cuts of $3 \arcsec - 12 \arcsec$ length perpendicular to the local orientation of the fibrils in this region using the CRisp SPectral EXplorer computer program \citep[CRISPEX; ][]{crispex, sstred}.

The space-time \CaII\ intensity plots of the cross-cuts reveal the oscillations of the bright fibrils \rf{(see panel (a) of Fig.~\ref{fig:cut_int} for an example). In order to bring out the oscillations more clearly, we processed the intensity data following the steps below, similar to the processes used by \citet{Kianfar20} and \citet{Jafarzadeh13,jafarzadeh17_2}:
\begin{enumerate}
 \item we integrated the intensity over the wavelength span that includes \CaIIK$_2$ peaks and the line core.
 \item we subtracted the wavelength-averaged wing intensity from the integrated core image to suppress the lower chromospheric contribution, in particular the 3-min acoustic oscillations. 
 \item we applied unsharp masking in the time direction using a median kernel of 3 pixels wide to lower residual image jitter.
 \end{enumerate}
Figure~\ref{fig:cut_int}a and \ref{fig:cut_int}b show the space-time intensity plots of an example cross-cut before and after the above enhancement procedure.}


\rf{We defined the trajectory of the POS oscillations in the processed space-time plots using a combination of manual and automatic approaches.
First, we manually defined a preliminary trail of each oscillation by choosing a sparse set of $(x,t)$ points along the oscillating fibril. Second, we linearly interpolated in time between these points to get a preliminary rough oscillation curve. Third, for each time step, similar to \citet{kuridze12}, we applied a Gaussian fit to the intensity at each time step around the preliminarily defined fibril location to get the precise location of the fibril. Finally, we smoothed the oscillation trajectories by applying a 1D boxcar average in time with a width of 64 seconds (8 time steps) to remove fluctuations on short timescales. We note that this smoothing introduces a bias in our sample by eliminating the oscillations with short periods. However, this does not affect our results as we are focusing on longer periods.}

\rf{Using the above procedure we created a sample of 605 POS oscillations of \CaIIK\ bright fibrils. The bottom panel of Fig.~\ref{fig:cut_int} shows the interpolated and smoothed trajectories of POS oscillations in the example cut marked in Fig.~\ref{fig:maps}f. The fitted fibril curves follow the unsmoothed fibrils in the background image rather well, which gives us confidence that the smoothing in time does not bias our results too strongly.}

We calibrated the overall axis of the oscillations by subtracting a linear trend from the trajectories so that the oscillation axis becomes parallel to the time axis \citep{jorrit15}. Then we computed the velocity of the POS oscillations by calculating the time-derivative of the amplitudes.

\subsection{Inversion and measurement of LOS oscillations}
\label{sb:inv}
We used the MPI-parallel STockholm inversion Code \citep[STiC;][]{jaime19} to derive the physical properties of the atmosphere from our observations. We used the inferred atmospheric velocity to estimate the LOS velocity in our fibril sample.

To solve the polarised radiative transfer equation, STiC uses the radiative transfer code RH \citep{RH} with Bezier solvers \citep{jaime2013}. STiC uses the equation-of-state that is acquired from the SME package functions \citep{2017A&A...597A..16P}, and includes partial redistribution (PRD) effects in the radiative transfer using the fast approximation method introduced in \citet{Jorrit2012-2}. STiC fits multiple spectral lines simultaneously for each individual pixel by assuming the 1.5D approximation (i.e. plane-parallel atmosphere). We ran STiC on the observations of the three \FeI , \CaIIIR , and \CaIIK\ lines to retrieve the physical properties in both photosphere and chromosphere. In our inversion runs, we assumed the statistical equilibrium and non-LTE condition for \CaII, including partial redistribution for the \CaIIK. For the \FeI\ line we assumed LTE. 

We did not aim to determine the magnetic field, and therefore only fitted Stokes $I$, even though our observations included all four Stokes parameters for \FeI\ and \CaIIIR. Because Zeeman broadening affects the width and shape of the line profiles, in particular for \FeI, we included the longitudinal magnetic field as a quantity in our atmospheric model with two nodes, and the transverse field with one node. 
Otherwise we used the same set-up and nodes to run the inversions as used in \citet{Kianfar20}, i.e., nine nodes in temperature, and four nodes for both $v_\mathrm{LOS}$ and $v_\mathrm{turb}$, located non-equidistantly in the optical depth span of $\logt = [-7,1]$. The inversion results are presented in Section~\ref{sb:inv_res}.

\section{Results}
\label{s:result}

\subsection{Inversion results}
\label{sb:inv_res}

\subsubsection{Region of interest}
\label{sbb:roi}

 \begin{figure*}
 \centering
 \includegraphics[width=\linewidth]{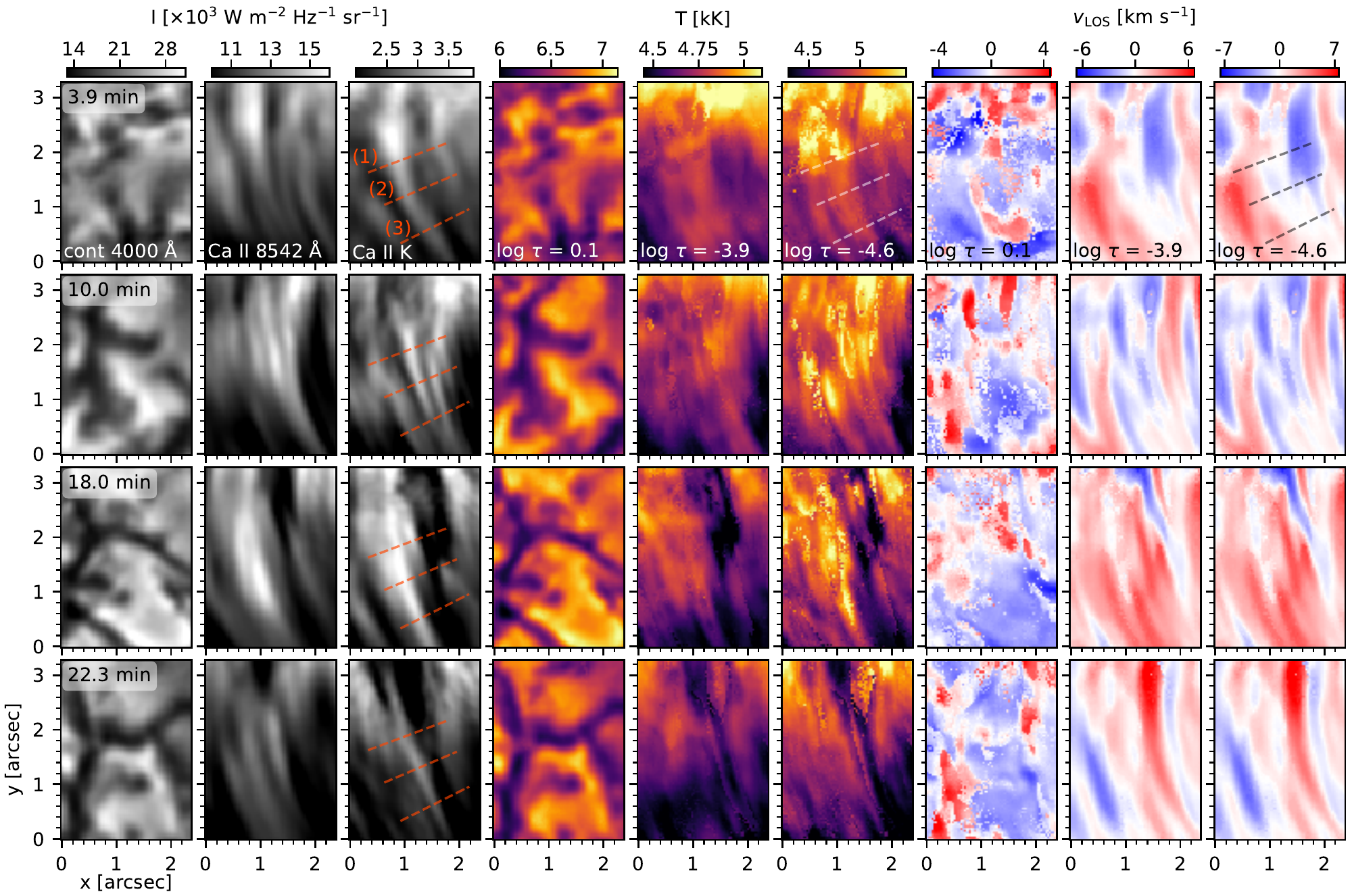}
 \caption{Time evolution of the physical properties of the RoI marked with a red box in Fig.~\ref{fig:maps}. \textit{Left panels:} intensity variation over time (from top to bottom) at the 4000~\AA\ continuum, and in the line cores of \CaIIIR\ and \CaIIK. \textit{Middle} and \textit{right panels:} time evolution of the temperature and \vlos\ at a photospheric depth ($\logt = 0.1$) and two different chromospheric depths, $\logt=-3.9$ and $-4.6$. Three cuts across the head, middle and the tail of this fibril are marked and numbered and their properties as function of time are displayed in Fig.~\ref{fig:3cut}. 
 }
 \label{fig:roi}
 \end{figure*}

To study the time-evolution of the fibrils, we first ran inversions on the time-series of a 2D region marked by the red box in Fig.~\ref{fig:maps}b. The results are shown in Fig.~\ref{fig:roi}. The region of interest (RoI) covers an area of $3\farcs1\times2\farcs2$ 
centering on a long-lived fibril (Fig.~\ref{fig:roi}, \CaIIK\ image at 22~minutes) with a clear transverse oscillation. The temperature and \vlos\ images in the photosphere (i.e., $\logt=0$) show a convection pattern consistent with the granulation pattern in the intensity images; there is higher temperature and upflows in the granules, and lower temperature and downflows in the intergranular lanes. The temperature in $\logt=-3.9$ at $t=10$~min and $t=18$~min almost follows the fibrillar patterns of the intensity images at \CaIIIR\ line-core. The fibrillar patterns in the temperature become more significant at $\logt = -4.6$ in the chromosphere where the bright fibrils appear as temperature enhancements in agreement with the \CaIIK\ intensity images \citep{Kianfar20}. The upflow and downflow structures in the \vlos\ images in the chromosphere do not particularly follow the fibrillar patterns, though they have the same average orientation as the fibrils \citep{Kianfar20}.

\begin{figure*}
 \centering
 \includegraphics[width=\linewidth]{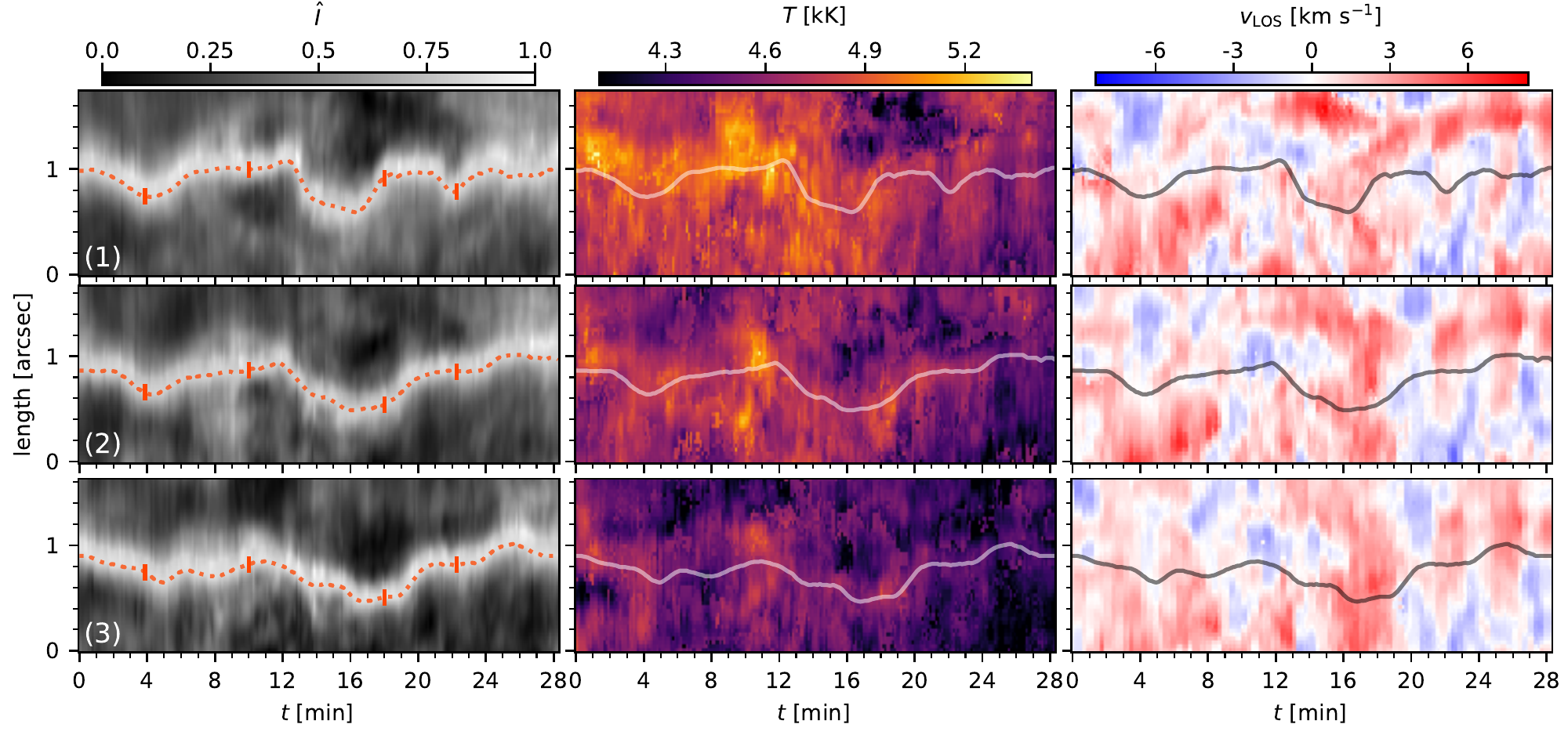}
 \caption{Physical properties of the cross-cuts marked in Fig.~\ref{fig:roi}. The \textit{left column} shows the intensity-over-time variations of the cuts at \CaIIK\ nominal line centre. The oscillation in the POS is marked by dotted red curves. The \textit{middle} and \textit{right columns} display the temperature and \vlos\ variations of the cuts as function of time in the chromosphere ($\logt = -4.31$). The POS oscillation is overplotted on these images as well. Length zero on the $y$-axis is where the cuts are marked by numbers in Fig.~\ref{fig:roi}.The $|$ symbols, marked on the curves in the left column, show the specific times of the fibril's evolution that are displayed in Fig.~\ref{fig:roi}.}
 \label{fig:3cut}
 \end{figure*}

We chose three perpendicular cuts (the dashed lines in Fig.~\ref{fig:roi}) across the central fibril in the RoI. Figure~\ref{fig:3cut} shows space-time plots of the intensity and inversion results along these cuts. The optical depth in which the inversion results are extracted is $\logt = -4.3$ because the \CaIIK\ line is most sensitive to the perturbations in the chromosphere at this optical depth \citep{Kianfar20}. The smoothed interpolated trajectories of the fibrillar POS oscillations in the cross-cuts are defined by the method explained in Sect.~\ref{sb:track} (overplotted curves in Fig.~\ref{fig:3cut}). They exhibit oscillations with a period $P$ of about 12~minutes that last for two periods. The period stays roughly constant along the fibril. We calculated the average phase speed for the oscillation along this fibril as 
\begin{equation}
\overline{v}_{\phi} = \frac{\Delta r}{\Delta t _{\phi_c}},
\label{eq:phase_vel}
\end{equation}
where $\Delta r$ is the average distance between the cuts and $\Delta t_{\phi_c}$ is the average time difference between the two points with the same phase in the oscillations. Accordingly the wave along this fibril propagates with an average speed of \rf{$\sim$12 km~s}$^{-1}$ from cut (1) to cut (3). The amplitude of the oscillations is highest (\rf{$A_{1} \approx 0\farcs4$} ) at the head of the fibril (i.e., close to the magnetic field concentration, top panels of Fig.~\ref{fig:3cut}) and it decreases to $A_{3} \approx 0\farcs2$ in the tail.

The temperature (middle column of Fig.~\ref{fig:3cut}) almost follows the oscillatory patterns in intensity with enhancements of 40--60~K. There are no obvious wave patterns in \vlos.


\begin{figure}
 \centering
 \includegraphics[width=\linewidth]{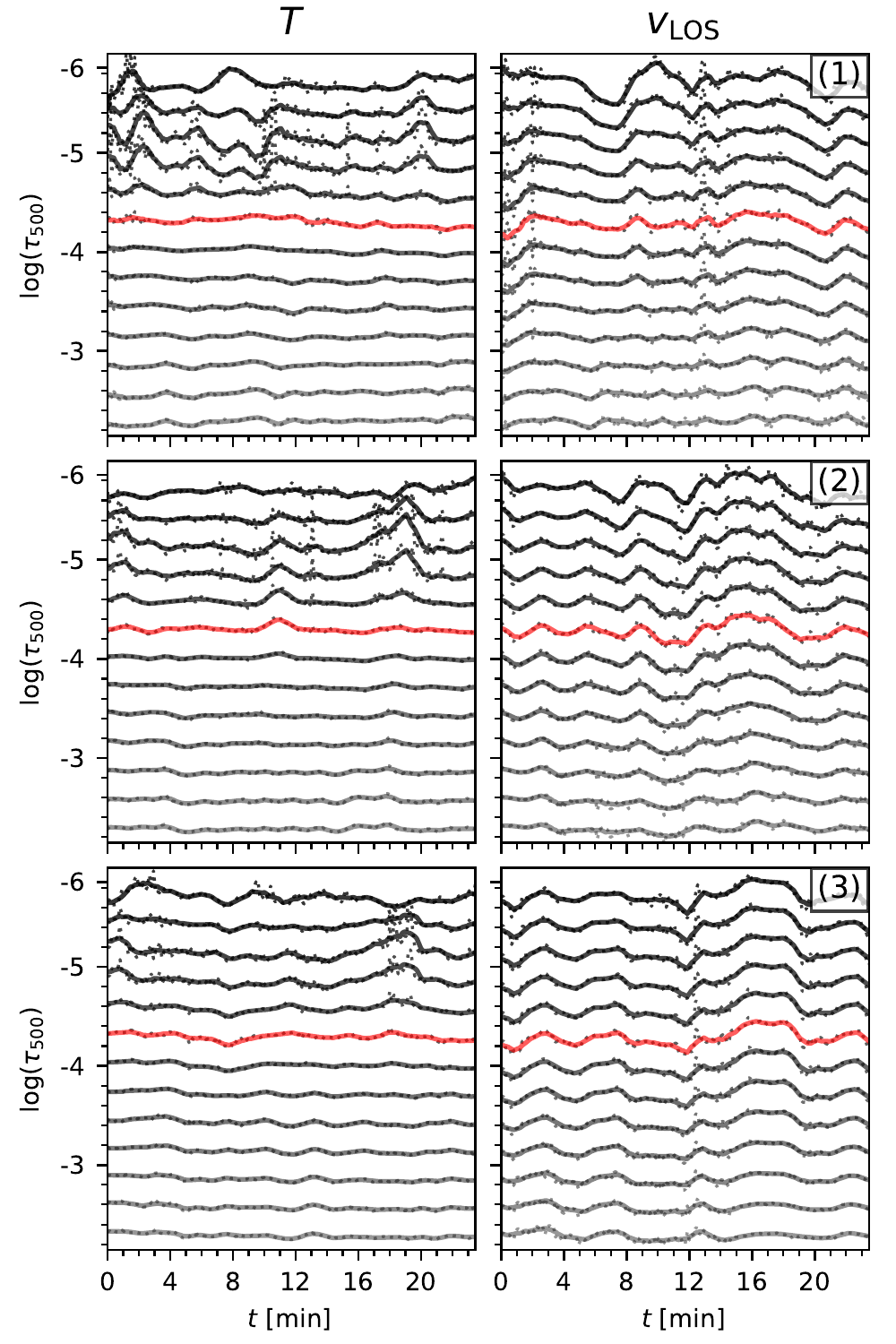}
 \caption{Temperature ({\it left column}) and \vlos\ ({\it right column}) variations of the three cuts marked in Fig.~\ref{fig:roi} over time. The panels show the temperature and \vlos\ variations along the POS oscillation trajectory (marked in Fig.~\ref{fig:3cut}) for a depth range where \CaIIK\ is sensitive to the perturbations in the atmosphere. Solid curves are the smoothed plots of the actual values that are shown with dots. Red-colored curves mark the depth where \CaIIK\ has the highest sensitivity to atmospheric perturbations. These curves are plotted individually in Fig.~\ref{fig:3cut_profile}.}
 \label{fig:seq}
 \end{figure}

We extracted the temperature and \vlos\ along the POS oscillations in the cuts of Fig.~\ref{fig:3cut} in the atmosphere between $\logt = [-6,-2]$. This is roughly the height range where \CaIIK\ line responds to perturbations of the physical properties in the atmosphere \citep{Kianfar20}. The results are shown in Fig.~\ref{fig:seq}. The oscillatory pattern in \vlos\ (right panel) starts around \rf{$\logt \sim -2.5$} and increases at larger heights, peaking between $\logt = [-3,-6]$. The temperature profiles (left panel) do not show any particular wave patterns. 

However, there are peaks forming at $t=11$ and 19~minutes, i.e, where the POS amplitude has extrema (see Fig.~\ref{fig:3cut}). These temperature peaks start appearing \rf{above $\logt = -4$} and get stronger up to $\logt=-5$. Higher up they get damped. This \rf{optical depth} range is where bright \CaIIK\ fibrils show temperature enhancements compared to their surroundings \citep{Kianfar20}.

\begin{figure}
 \centering
 \includegraphics[width=\linewidth]{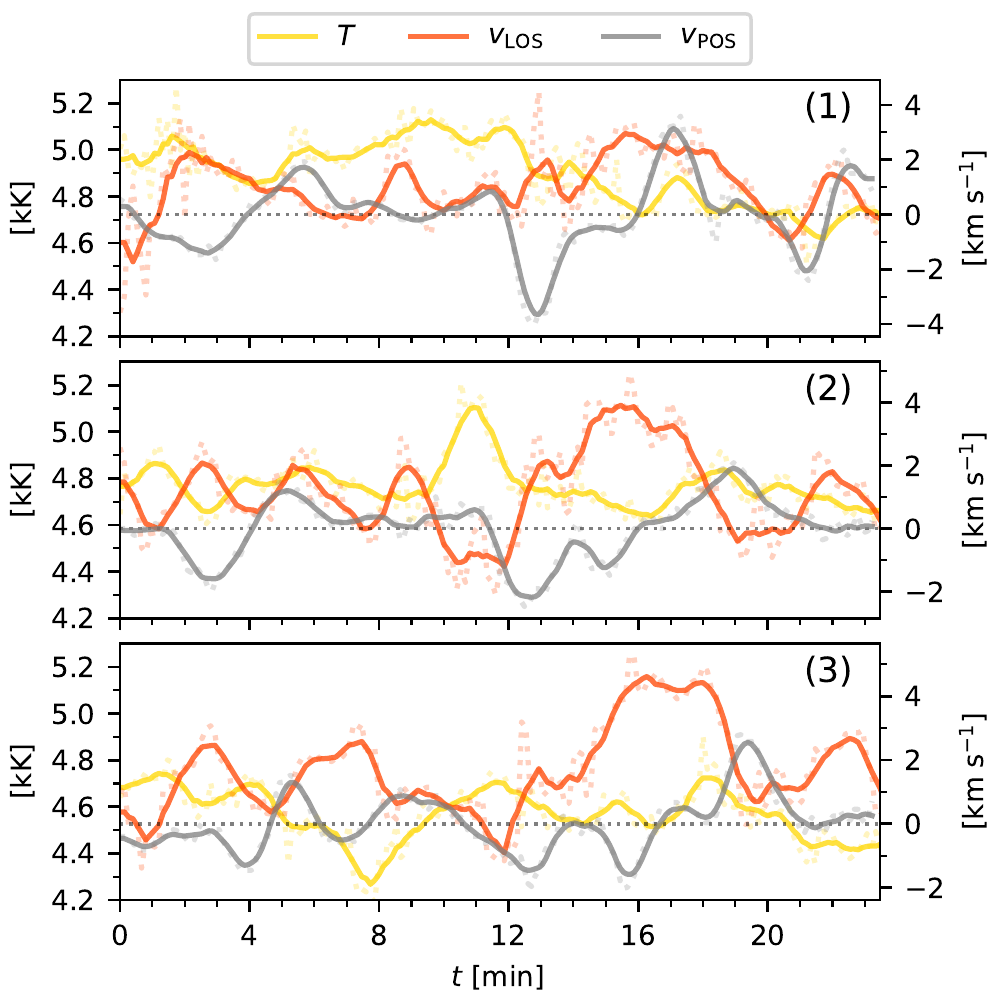}
 \caption{Variations of the temperature (\textit{left column}) and \vlos\ (\textit{right column}) of the perpendicular cross-cuts shown in Fig.~\ref{fig:roi}. They are plotted based on their values extracted along the oscillation trajectory in POS marked in Fig.~\ref{fig:3cut} at the depth $\logt = -4.31$ (i.e., where the \CaIIK\ line is most sensitive to atmospheric perturbations). The oscillation of \vpos\ is plotted in grey for comparison.}
 \label{fig:3cut_profile}
 \end{figure}

While the peaks in the temperature profiles get smaller as we move towards the tail of the fibril (i.e., from cut~1 to cut~3), the amplitude of \vlos\ oscillations increases. Figure~\ref{fig:3cut_profile} shows a comparison of the time evolution of $T$ and \vlos\ at $\logt = -4.31$ to \vpos. It shows that 
both the temperature variation and its average decrease as we move from the head to the tail of the fibril. The amplitude of \vpos\ behaves the same. The \vlos\ curves show different behaviour, with the highest amplitude peak at about \rf{$t=16$}~min in cut~(3), i.e., close to the tail of the fibril. We further discuss the \vpos\ and \vlos\ oscillations of our sample and their wave properties in Sect.~\ref{sb:oscillations}.

\subsubsection{Cross-cuts} \label{sbb:crosscut}

 \begin{figure*}
 \centering
 \includegraphics[width=\linewidth]{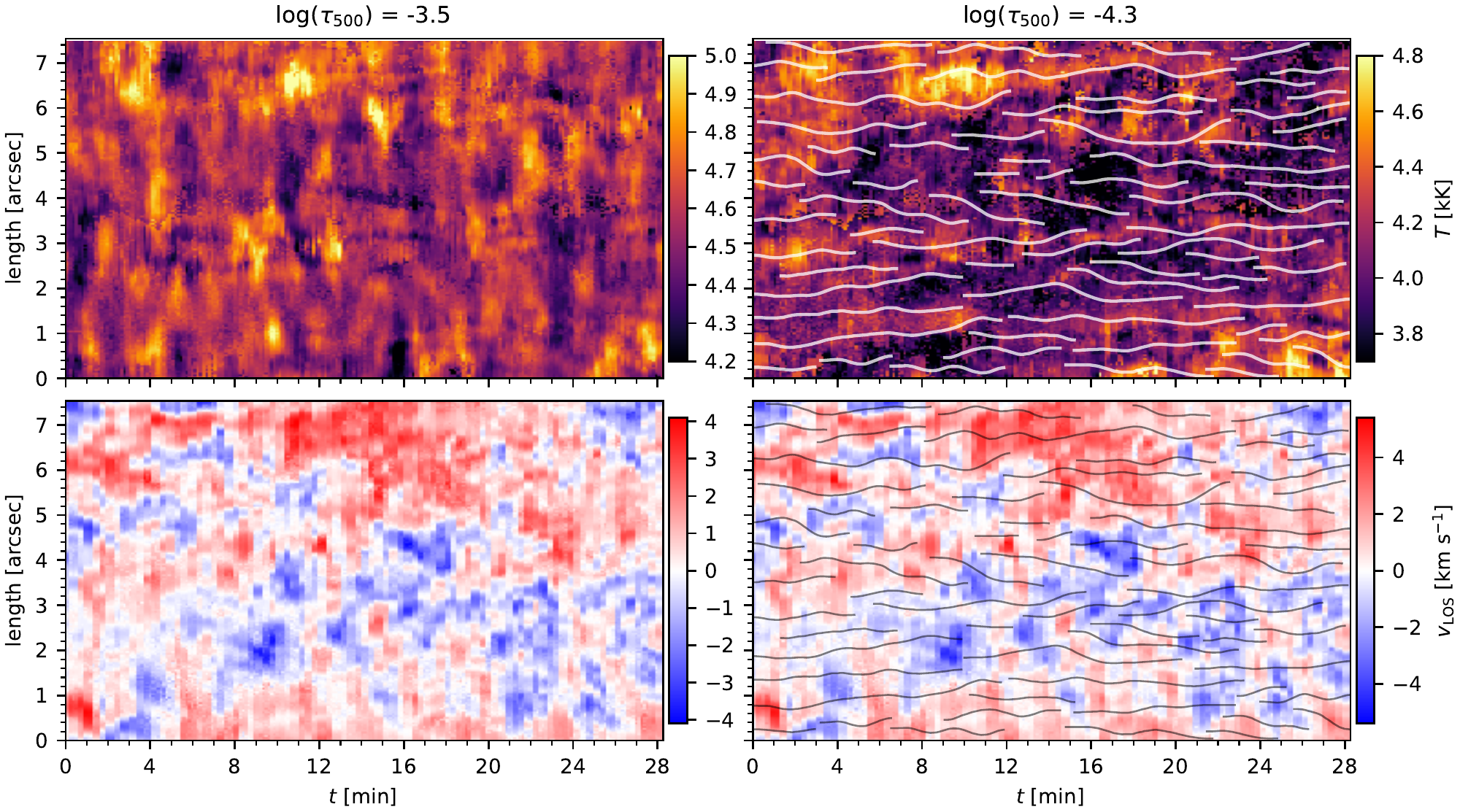}
 \caption{Time evolution of the physical properties of the example cross-cut shown in Fig.~\ref{fig:cut_int}. The \textit{left column} shows the inversion results in the lower chromosphere and the \textit{right column} represents the results higher in the chromosphere. The \textit{top panels} show the temperature of the cut in the chromosphere. The \textit{bottom panels} show space-time images of the LOS velocity. The POS oscillations are overplotted on the \textit{right panels}. The zero point of the cut length is marked with a white dot in Fig.~\ref{fig:maps}.}
 \label{fig:cut_inv}
 \end{figure*}

Figure~\ref{fig:cut_inv} shows the temperature and \vlos\ of the example cut shown in Fig.~\ref{fig:cut_int} at two chromospheric heights. The temperature in the lower chromosphere ($\log \tau_{500} =-3.5$) demonstrates a very faint pattern of fibrillar structures. Higher up, at $\log \tau_{500}=-4.31$, \rf{the temperature image shows more recognizable fibrillar structures that resemble similar patterns to the space-time intensity image} (Fig.~\ref{fig:cut_int}) with enhanced temperature at the location of the bright fibrils \citep{Kianfar20}. The \vlos\ space-time images do not demonstrate a clear correlation to the fibrillar oscillation trajectories in the POS and appears similar at both heights, with only a slight difference in the range of the velocity values.

\subsection{3D oscillations} \label{sb:oscillations}

 \begin{figure*}
 \centering
 \includegraphics[width=0.97\linewidth]{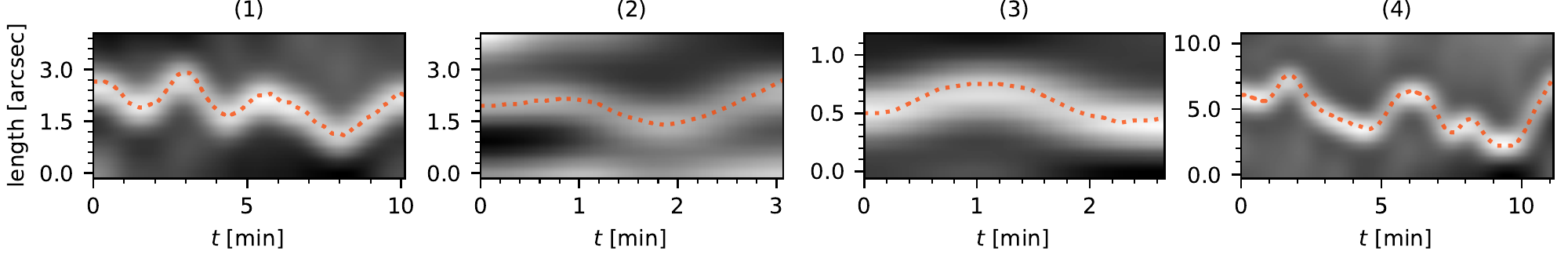}
 \includegraphics[width=0.495\linewidth]{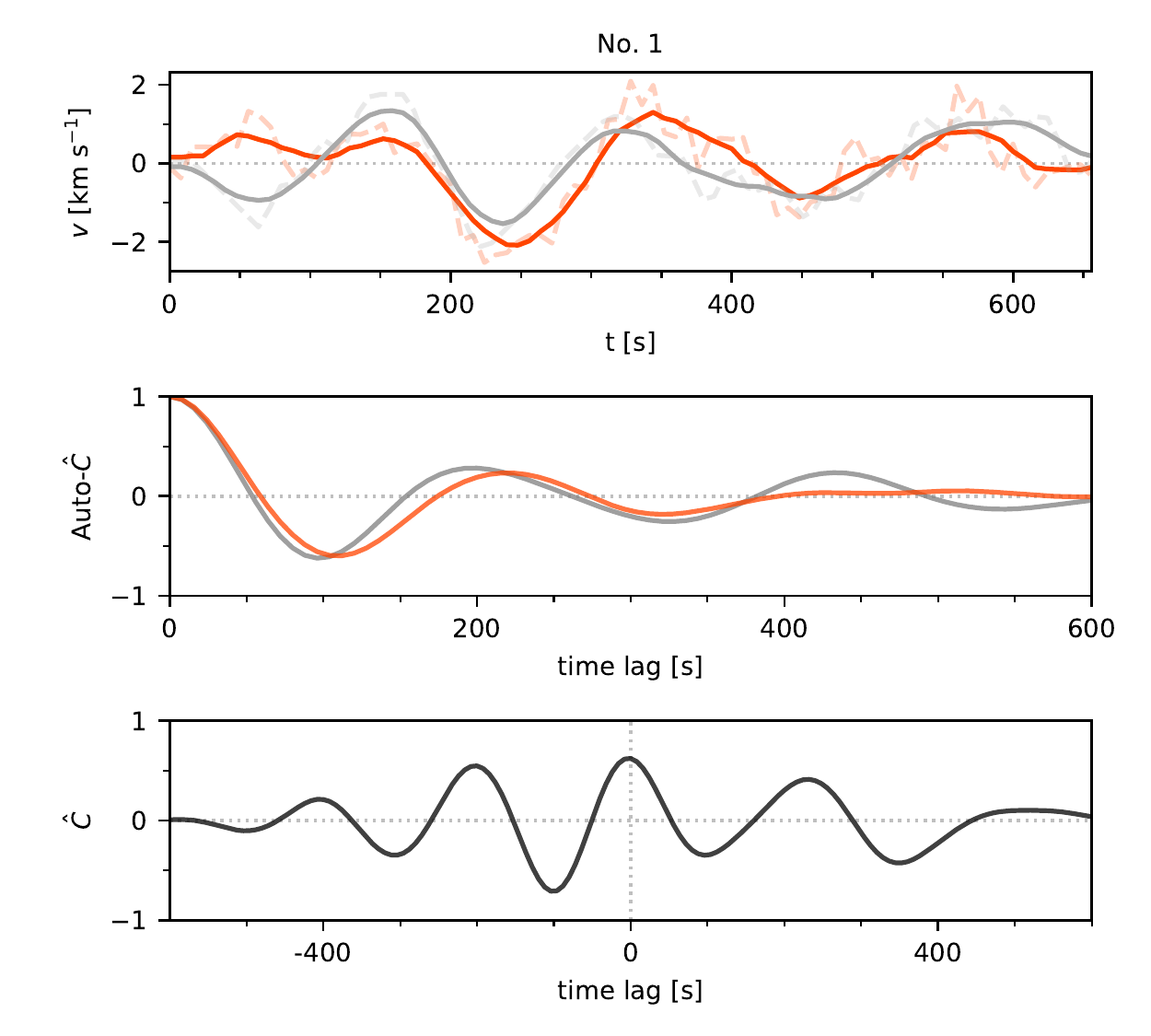}
 \includegraphics[width=0.495\linewidth]{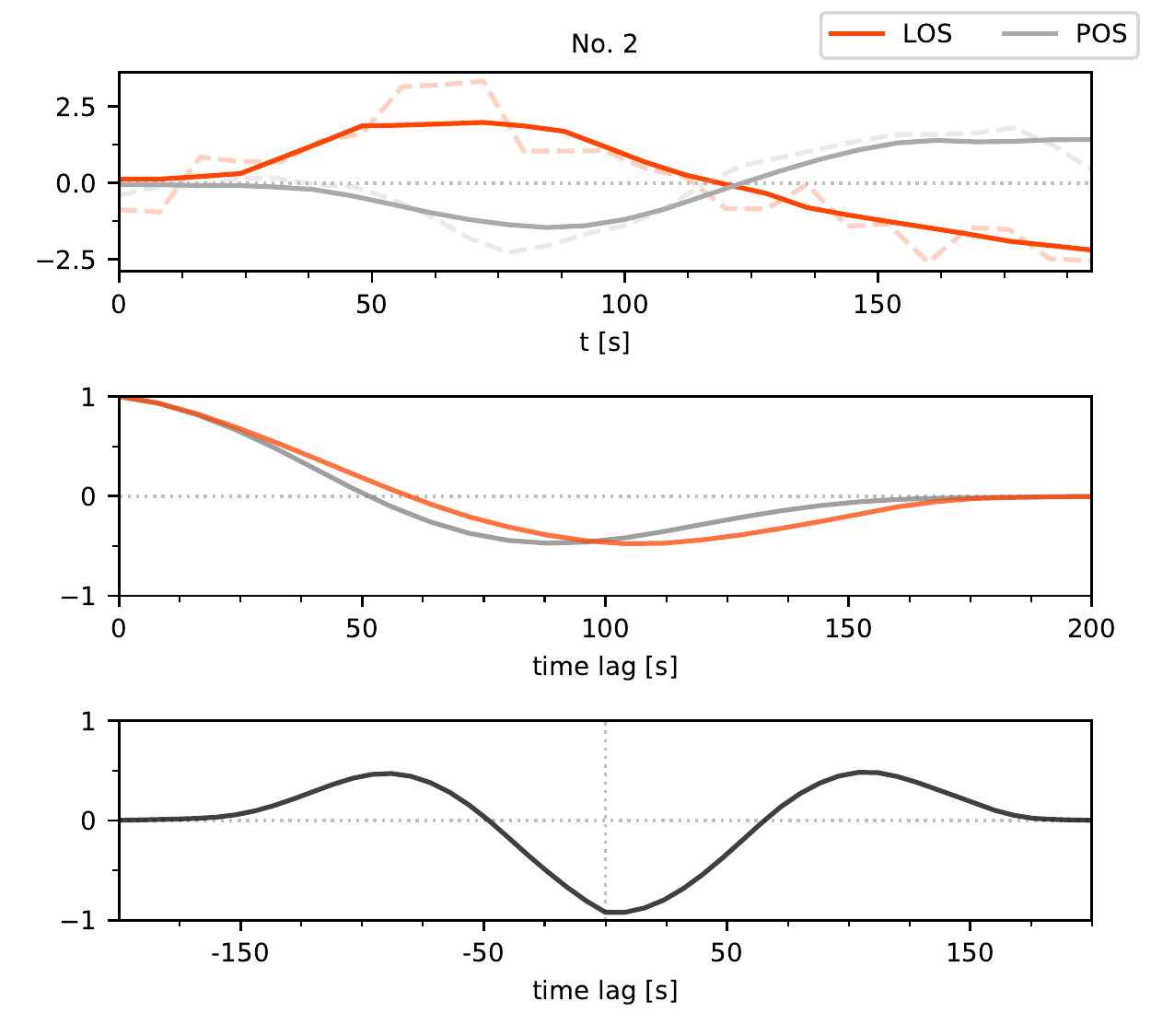}
 \includegraphics[width=0.495\linewidth]{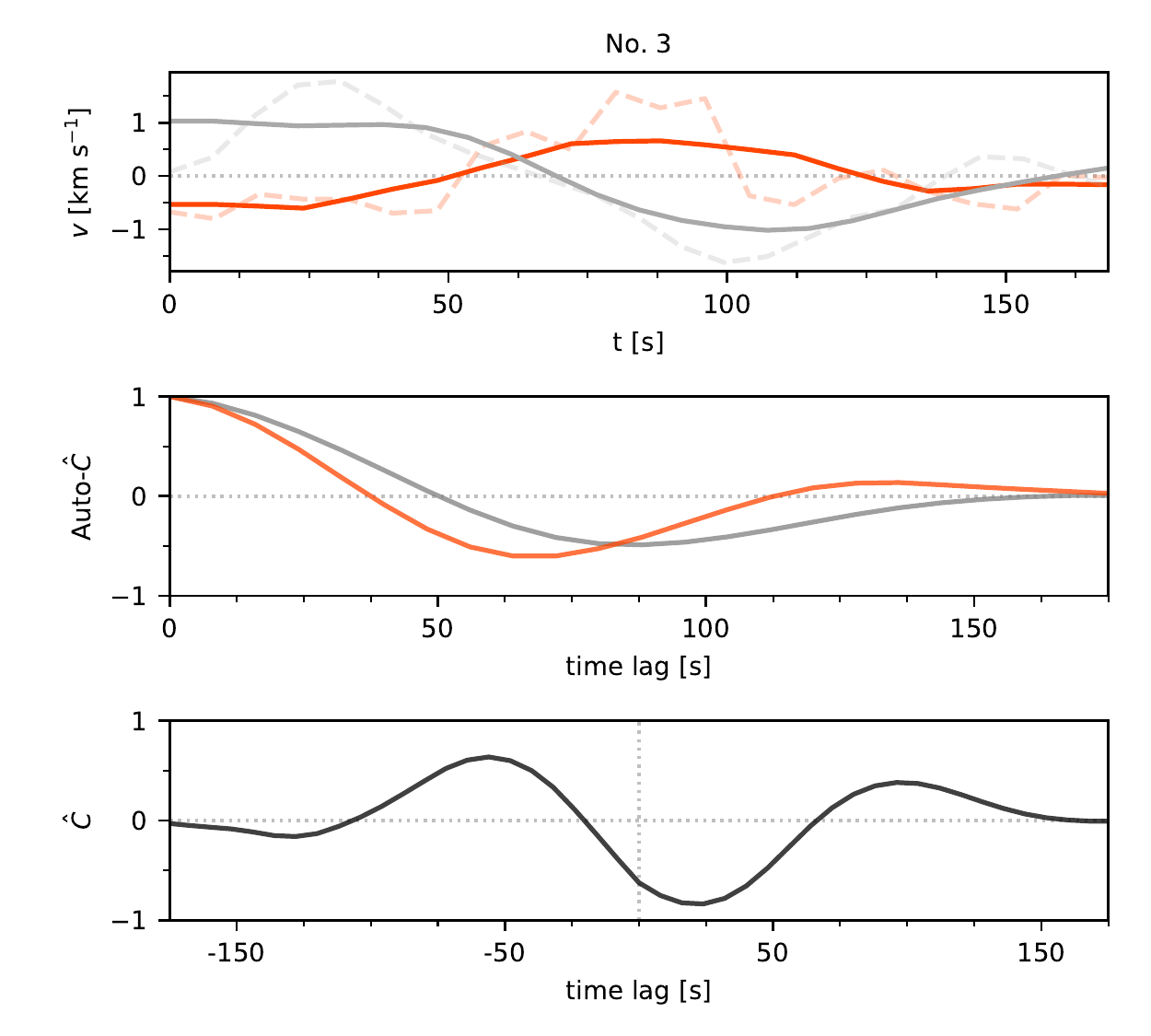}
 \includegraphics[width=0.495\linewidth]{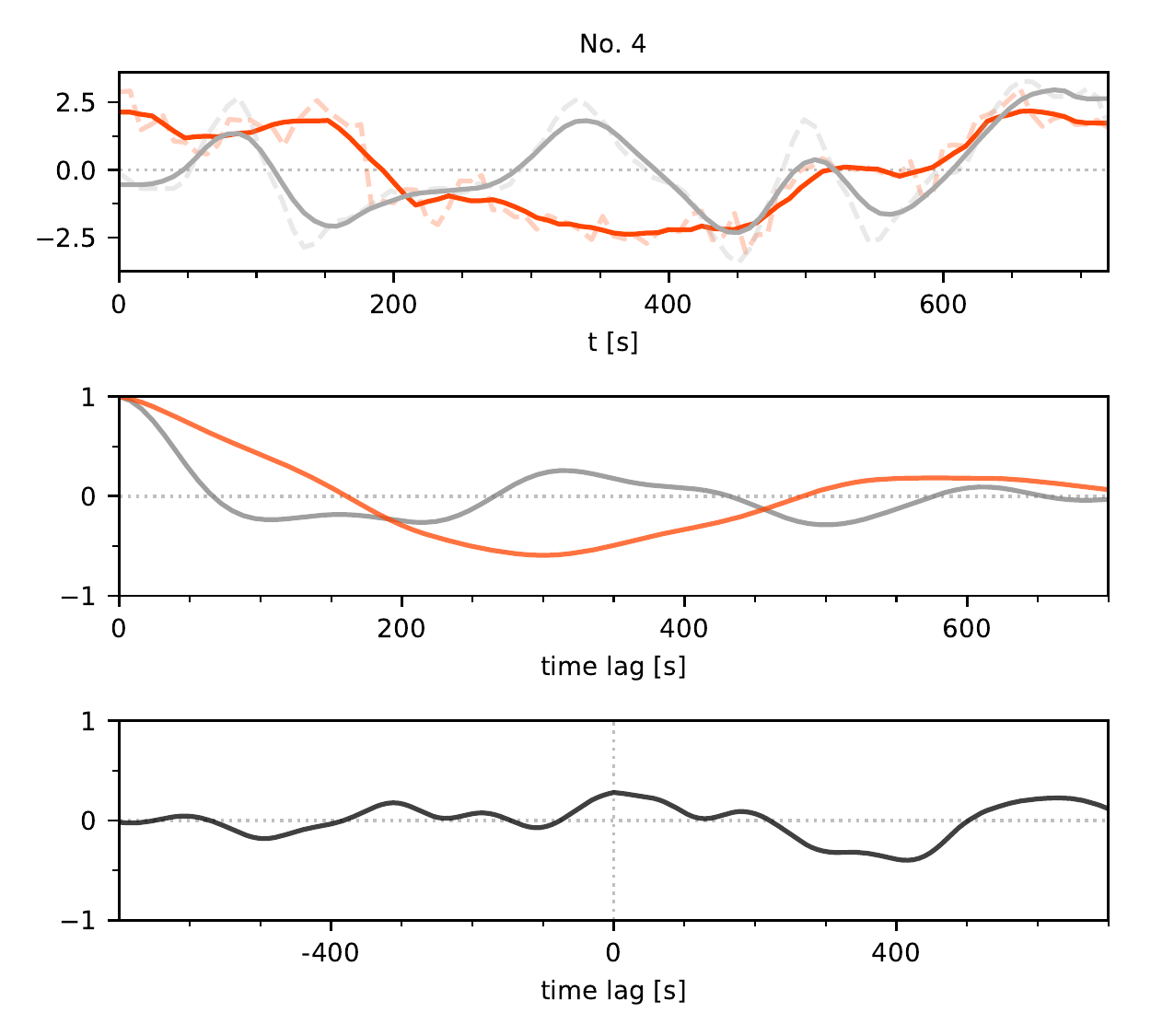}
 \caption{Comparison of the velocity oscillations of the four examples marked with numbers in Fig.~\ref{fig:cut_int}. These examples represent four different groups of oscillations in our sample of \CaIIK{} fibrillar oscillations. 
The top row shows time-slices of the example fibrils, with the smoothed fitted POS motion overplotted with red dotted curves. Below, we show three plots for each of the four fibrils.
 The \textit{top plot} shows the POS velocity (red), and the LOS velocity (grey) with a solid curve for the smoothed data and a dotted curve for the unsmoothed data. The \textit{middle plot} shows the auto-correlation function of the POS (grey) and the LOS (red) velocity. The {\it bottom plot} shows the normalised cross-correlation function ($\hat{C}$) of the POS and LOS velocity. For the interpretation of the above plots and more details on the different fibrillar groups see Sect.~\ref{ssb:osc_prop} and \ref{sbb:stats}. An animation of this figure is available online that shows all \rf{468} fibrils in the subsample.}
 \label{fig:cut_cat}
 \end{figure*}

From the POS oscillation trajectories (Sect.~\ref{sb:track}) and space-time inversion results of the cuts across the fibrils (Sect.~\ref{sbb:crosscut}), we determined the LOS velocity along the fibrils. To do so, we extracted the \vlos\ values along the smoothed trajectory of the POS oscillations (Sect.~\ref{sb:track}) at $\logt = -4.31$, i.e., the height where temperature and velocity at fibrillar locations are most sensitive to the perturbations in the chromosphere \citep{Kianfar20}. In addition, we computed the velocity in the POS by taking the time derivative of the POS displacement. The top panels of Fig.~\ref{fig:cut_cat} show \vlos\ an \vpos\ along the four example fibrils marked with numbers in Fig.~\ref{fig:cut_int}. 

\subsubsection{Determination of the oscillation properties} \label{ssb:osc_prop}

We derived the wave properties in our oscillation sample through a combination of automatic and manual approaches. First, we computed the autocorrelation of each \vlos\ and \vpos\ curve. Besides the peak at a time lag $\Delta t=0$~s, a curve with an oscillation shows a secondary peak at a time lag roughly equal to the dominant period of the oscillation. Because the velocity curves contain noise, and might contain partial oscillations, interference patterns or oscillation with changing period, the secondary peak can have a much lower amplitude. Combining the autocorrelations with direct visual inspection of the curves, we found a subsample of 468 fibrils out of the total sample size of 605 that have a clear oscillation in both \vlos\ and \vpos. 
The middle panels of Fig.~\ref{fig:cut_cat} show the autocorrelations for the four example fibrils marked with numbers in Fig.~\ref{fig:cut_int} that belong to this category.

We then derived periods for this subsample by measuring the time of each extremum, and compute "partial periods" as the time difference between two consecutive minima or maxima. As the periods are not constant throughout one oscillation, we assigned an average and a dominant period ($\bar{P}$, $P_\mathrm{dom}$) as the mean and maximum of the partial periods for \vlos\ and \vpos\ in each fibril. We also measured the amplitude of each extremum, and used that to assign an average amplitude $\bar{A}$ for \vlos\ and \vpos\ \rf{for each oscillation}.

To quantify the correlation between \vlos\ and \vpos, we calculated the normalised cross-correlation function, $\hat{C}(t)$, for each fibril. \rf{Cross-correlation curves of the example} fibrils are shown in the bottom panels of Fig.~\ref{fig:cut_cat}. \rf{We considered fibrils with $max\bigl\{|\hat{C}(t)|\bigl\}~> 0.5$ as potentially exhibiting correlated motion. Then we visually inspected those fibrils to confirm that \vlos\ and \vpos~indeed have similar periods and oscillation patterns. Those fibrils with a sufficiently high $max \bigl \{|\hat{C}(t)| \bigl\}$ and a difference in period between the two velocity components smaller than 20\% defined a subset of 373 fibrils out of the \rf{468} oscillations.}

For this subset we computed the phase difference between the oscillations as 
\begin{equation}
\Delta \phi = \pi \frac{\Delta t}{P_M}, \quad -\pi \le \Delta\phi \le \pi,
\label{eq:dphi}
\end{equation}
where $P_M = (\bar{P}_\mathrm{LOS} + \bar{P}_\mathrm{POS} )/2$. The online animation accompanying Fig.~\ref{fig:cut_cat} shows all fibrils in our subset.

\subsubsection{Statistical analysis}
\label{sbb:stats}

 \begin{figure*}
 \centering
 \includegraphics[width=\columnwidth]{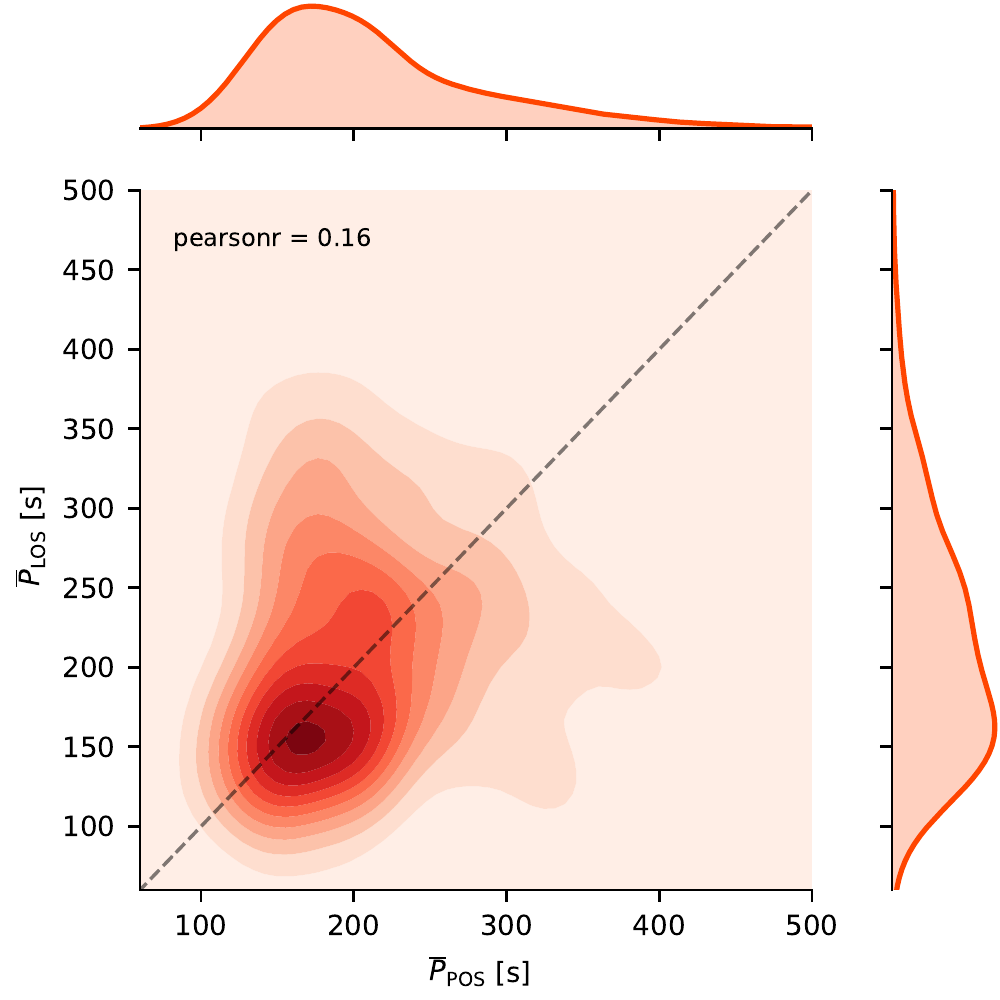}
 \includegraphics[width=\columnwidth]{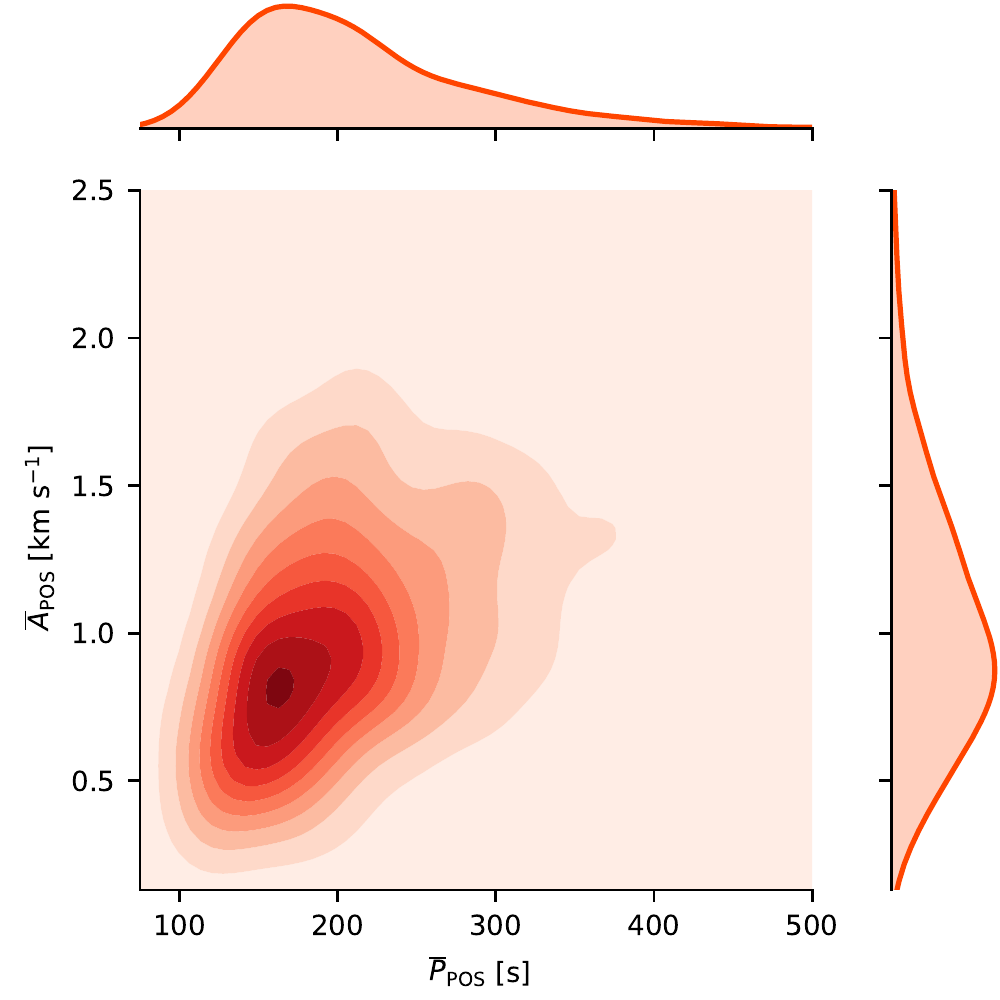}
 \includegraphics[width=\columnwidth]{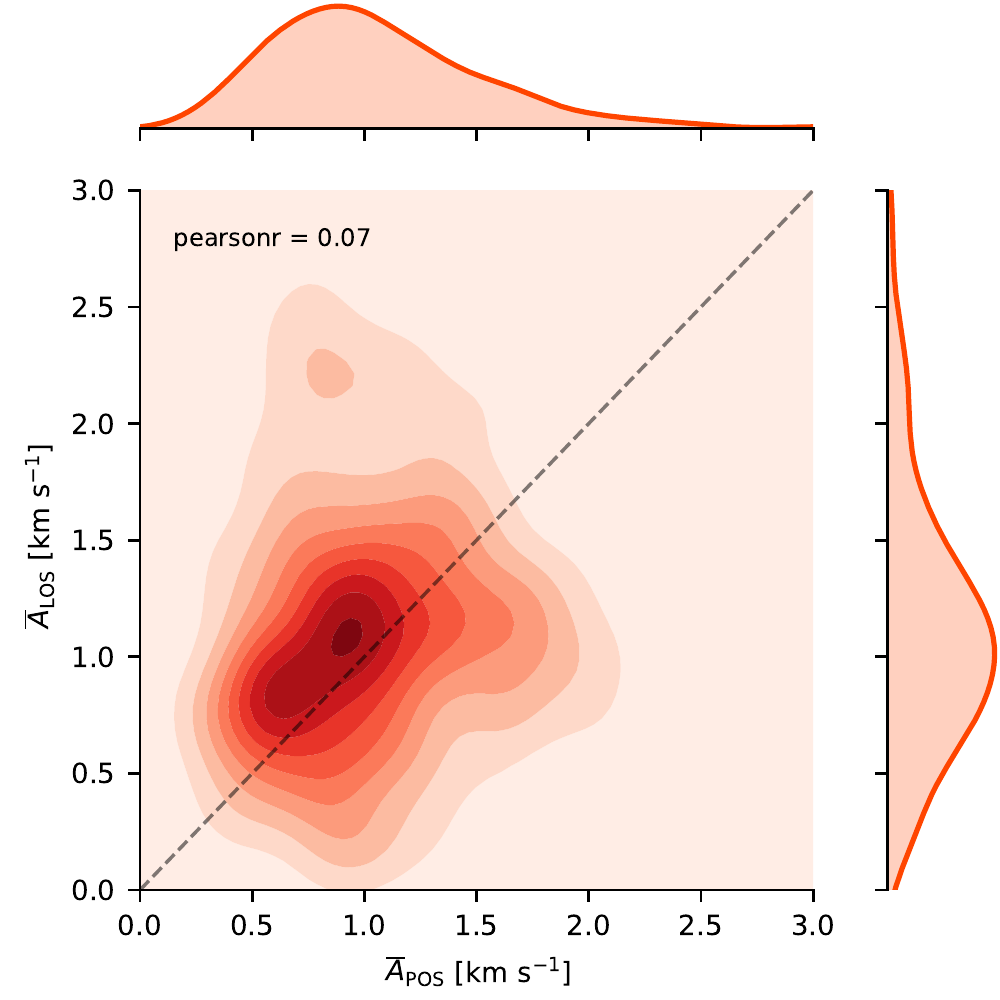}
 \includegraphics[width=\columnwidth]{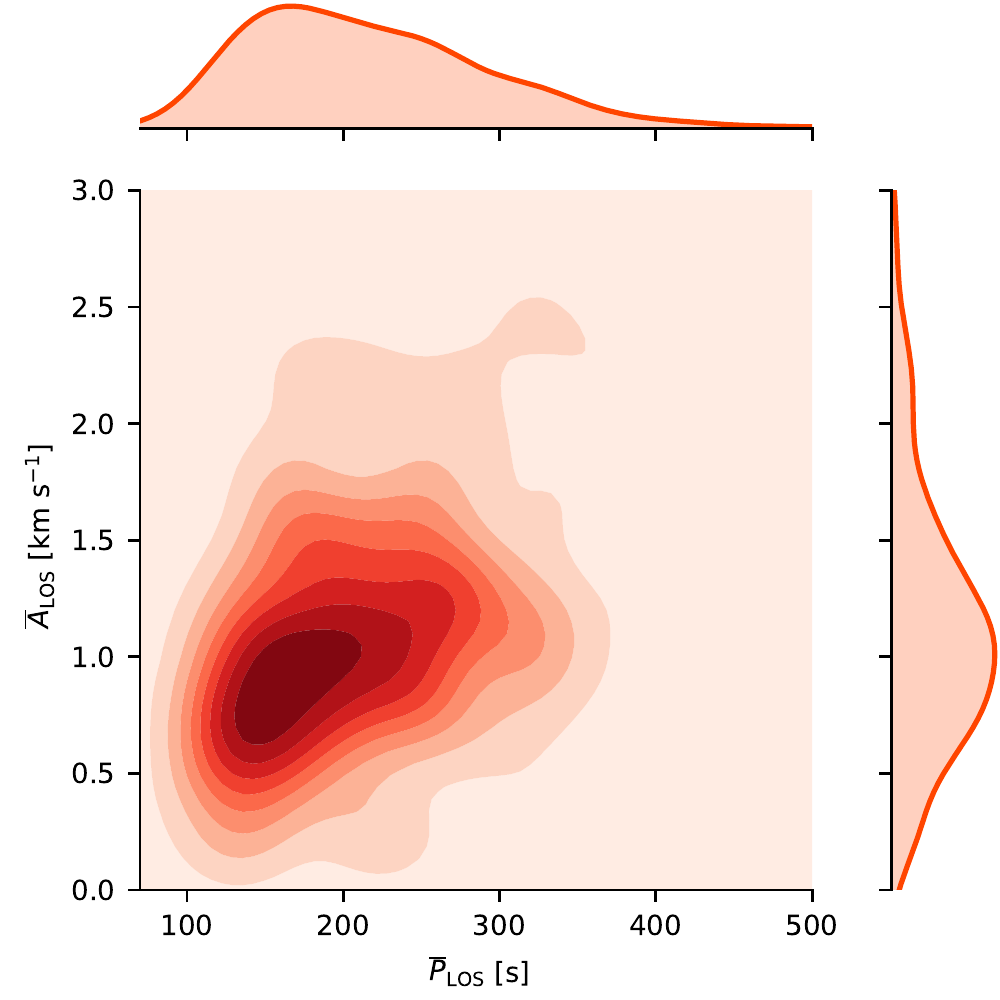}
 \caption{Kernel density estimate (KDE) plots and histograms of the average period $\bar{P}$ and average amplitude $\bar{A}$ of the \rf{468} fibrils in our sample that show POS and LOS oscillations. The panels display the quantities indicated in the axis labels. The histogram distribution for each quantity is shown in the side-plots corresponding to their axis. The Pearson correlation coefficient $r$ is given in the corner of the left-hand panels.}
 \label{fig:A_P}
 \end{figure*}

 \begin{table}
\caption{\rf{Statistic summary of the fibril oscillations along LOS and POS directions.}} 
\label{table:statistics} 
\begin{tabular}{l r r r r r} 
\hline\hline 
 & min & max & mean & median & std. dev.\\
\hline 
$\overline{P}_{\rm{POS}}$ (s) &92 & 662 & 207 & 190 & 74 \\
 $\overline{P}_{\rm{LOS}}$ (s) &76 & 584 & 219 & 205 & 78\\
 $\overline{A}_{\rm{POS}}$ (\kms) &0.21 & 3.05 & 1.03 & 0.96 & 0.45 \\
 $\overline{A}_{\rm{LOS}}$ (\kms) &0.09 & 4.19 & 1.16 & 1.06 & 0.62\\
 

\hline 
\end{tabular}
\end{table}
 
Figure~\ref{fig:A_P} shows the distribution of and correlations between the measured periods and amplitudes. Table~\ref{table:statistics} shows their minimum maximum, average, and standard deviation.

The top-left panel of Fig.~\ref{fig:A_P} shows a weak but positive correlation between $\overline{P}_{\rm{POS}}$ and $\overline{P}_{\rm{LOS}}$ especially between periods of \rf{130 to 210~s}. In contrast, the amplitudes (lower-left panel) do not appear to have a correlation between the two POS and LOS directions. In general the \vlos\ oscillations tend to have \rf{slightly larger} amplitudes compared to \vpos. This is also confirmed by the mean values in Table.~\ref{table:statistics}. There appears to be no correlation between the amplitude and period in the POS direction (upper right), and likewise for the LOS direction (lower right).

 \begin{figure}
 \centering
 \includegraphics[width=\linewidth]{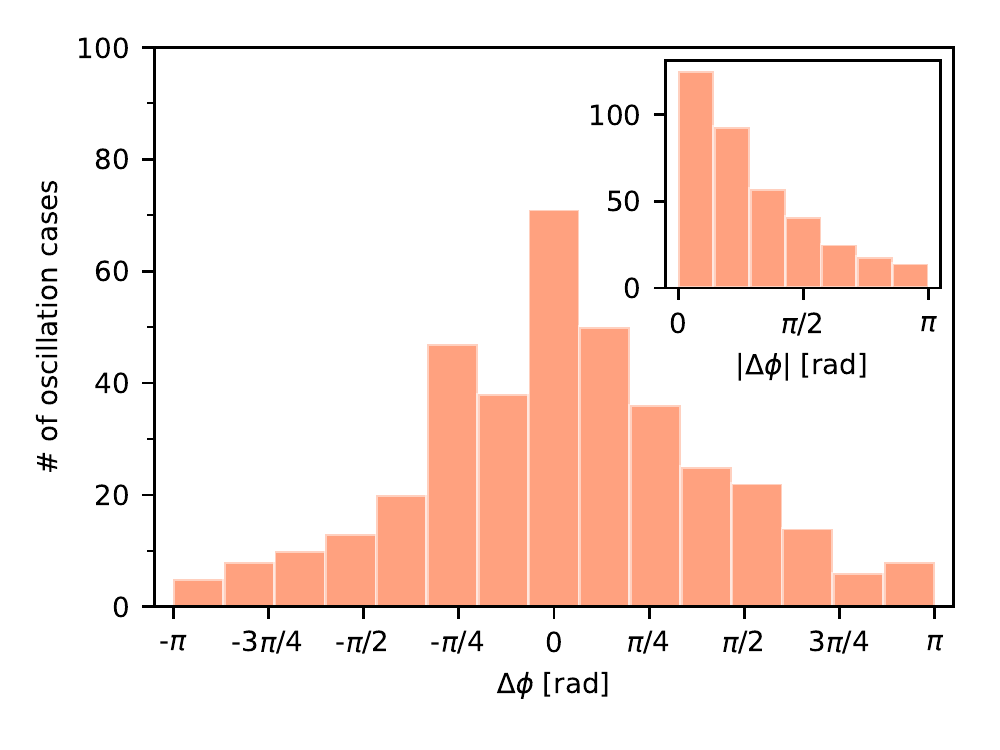}
 \caption{Distribution of the phase difference $\Delta\phi$ and $|\Delta\phi|$ for the \rf{373} fibrils in our sample that show velocity oscillations in the POS and LOS direction with similar periods.}
 \label{fig:dphi}
 \end{figure}

We show the distribution of the phase difference between the \vpos\ and \vlos\ oscillations for the \rf{373} fibrils in our sample that indicate correlated velocity oscillations in the POS and LOS directions in Figure~\ref{fig:dphi}. The observed phase difference spans the whole range of [$-\pi,\pi$]. This implies that the total velocity vectors $\mathbf{v}_{tot} = ({v}_{\rm{POS}},v_{\rm{LOS}})$ can range from fully linearly polarized ($\Delta \phi = 0$, $-\pi$, or $\pi$) to fully circularly polarized ($\Delta \phi = \pm \pi/2$). 

We also show the distribution of $|\Delta \phi|$ in Fig.~\ref{fig:dphi}. Because our sample size is limited and we are only interested in absolute phase differences, this lowers the statistical noise in our sample. There is a clear peak in the distribution for linearly polarized waves.

\begin{figure}
 \centering
 \includegraphics[width=\linewidth]{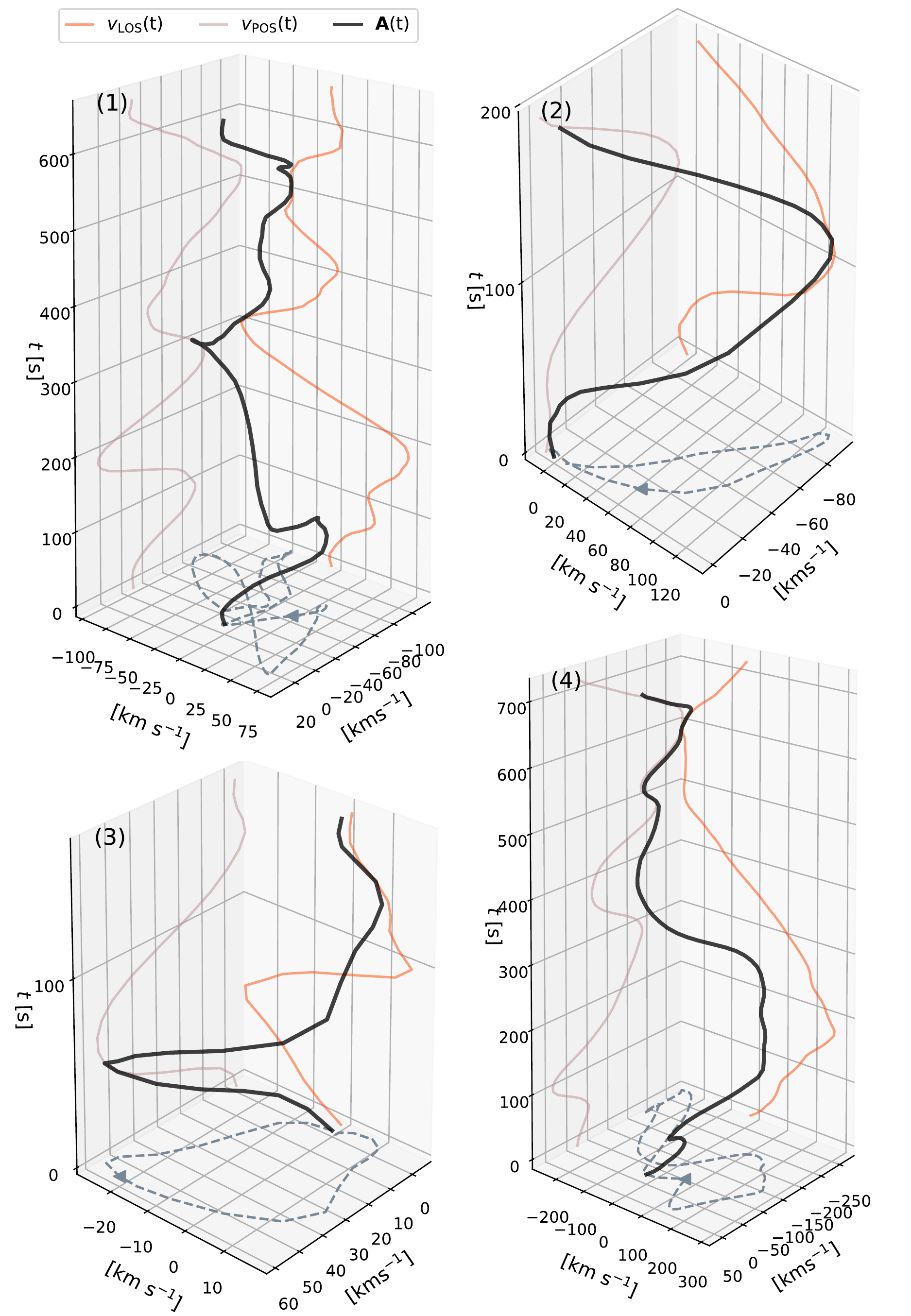}
 \caption{Time evolution of the motion of a point along the fibril axis $\mathbf{A}(t)$ of the four example oscillations shown in Figs.~\ref{fig:cut_int} and~\ref{fig:cut_cat}. The oscillations in the POS and LOS are shown in the side planes and the projection of the total motion are plotted in the plane perpendicular to the POS and LOS. The small arrow shows the direction of the motion in the bottom plane. An animated version of this figure showing all \rf{468} correlated fibrils in our sample is available online.}
 \label{fig:3d_spiral}
 \end{figure}

In order to visualize the motion of the fibril in 3D we integrated the LOS velocity to obtain the LOS displacement as a function of time. Together with the measured POS amplitude we can then trace the motion of a particular point along a fibril as a function of time. We display these curves in Fig.~\ref{fig:3d_spiral} for the four example fibrils shown in Figs.~\ref{fig:cut_int} and~\ref{fig:cut_cat}.

The phase differences between the velocity oscillations of the examples~(1) and~(2) are 0 and $\pi$ as shown in Fig.~\ref{fig:cut_cat}, suggesting a linear polarized wave. This linear behaviour is seen roughly in the total motion of the example~(2) in Fig.~\ref{fig:3d_spiral} but not so clearly in example~(1). Example~(3) has a $\Delta \phi \approx \pi/2$, and therefore the projection of the amplitude vector traces out an elliptical path. Example~(4) has velocities that appeared to have low-amplitude short-period velocities with different periods in the LOS and POS directions. However, the projection of the amplitude vector reveals it carried a long-period linear oscillation as well as the short period oscillations.

\begin{figure}
 \centering
 \includegraphics[width=\linewidth]{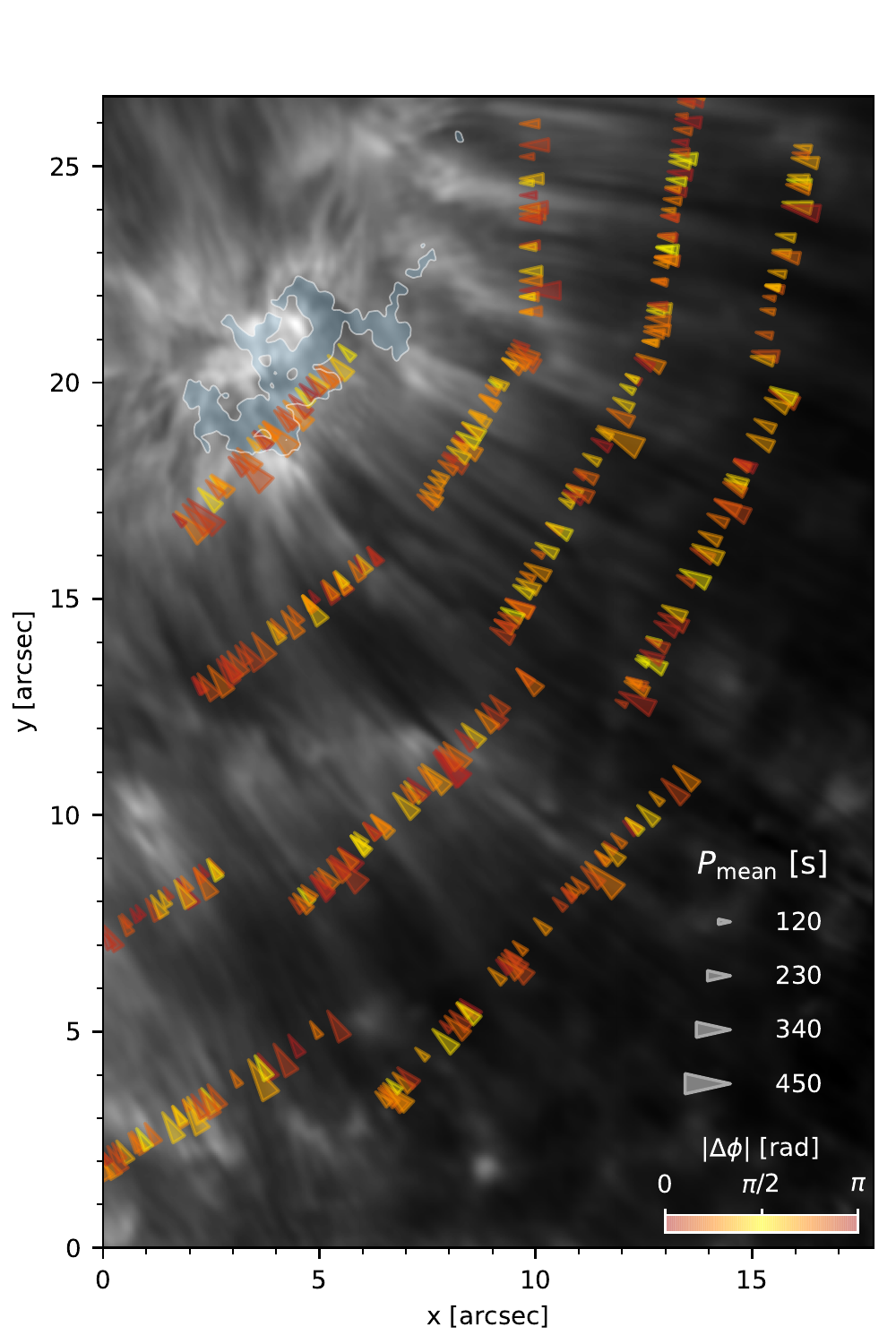}
 \caption{Spatial distribution of measured fibril oscillation periods and LOS-POS phase differences.
 Each symbol marks the coordinates of the oscillation point. The $\blacktriangle$ symbols mark the points where an oscillation is detected in both \vpos\ and \vlos. The size of the $\blacktriangle$ symbols is proportional to their average period. The color-coding of yellow, orange and red represents the absolute values of the phase difference between the \vlos\ and \vpos\ oscillations where their period is correlated. The coordinates of the oscillations that do not have a clear correlation between their $\overline{P}_{\rm{POS}}$ and $\overline{P}_{\rm{LOS}}$ are shown with gray triangles. The orientation of the $\blacktriangle$ symbols shows the perpendicular direction to the cuts (Fig.~\ref{fig:maps}-f) across the bright fibrils. The cross symbols mark the points where an oscillation is detected in the POS direction but not in the LOS direction. The white filled contour lines indicate areas in the photosphere with magnetic concentrations, where the footpoints of the fibrils are located.}
 \label{fig:scatter}
\end{figure}


Finally, we investigated the spatial distribution of the periods and phase differences. Our results are shown in Fig.~\ref{fig:scatter}. There is no correlation between location of the fibril cross-cuts and their periods and phase difference.


\paragraph{Summary of results} We summarise our main findings: we selected a number of cross-cuts through \CaIIK\ bright fibrils, and defined a sample of 605 fibrils along these cross-cuts. We determined the velocity amplitudes and periods of the fibril oscillations for both the LOS and POS directions. \rf{The distribution of periods and amplitudes in both directions is similar, with a mean period of around $2.2\times10^2$~s, and a mean velocity amplitude of 1.1~\kms.}

Because the sample was selected based on POS oscillations, all of them exhibit a POS oscillation. \rf{The majority of them (77\%, i.e., 468 fibrils out of 605 oscillations in total) show a clear LOS oscillation pattern. Of those, 373 fibrils (80\% of the subsample) showed oscillations with similar periods in both the LOS and POS directions. For these, we measured the phase difference between the two oscillation directions. All phase differences occur, but there is a preference for zero phase difference.}
\section{Discussion}

The periods and POS velocity amplitudes that we measure here fall in the range of transverse oscillations observed in H$\alpha$ fibrils as reported by \citet{2013ApJ...768...17M}. The oscillations in H$\alpha$ spicule-like events reported by \citet{2021ApJ...921...30S} have smaller periods and larger amplitudes.

The median period in POS oscillations of about \rf{190}~s that we measure is larger than the median period of 84~s reported in
\citet{jafarzadeh17_2} for \CaIIH\ fibrils observed with a 0.11~nm filter. Table~2 of \citet{jafarzadeh17_2} lists observed periods and amplitudes of transverse oscillations in fibrils, mottles, and spicules observed in H$\alpha$ and \CaIIH\ in different studies using a variety of instruments. Specifically, the observed periods ranged between 16~s and 600~s, roughly consistent for all studies. However, the differences in periods and amplitudes might be caused by multiple reasons, such as differences in the observed target, observing cadence, procedures used to define and detect spicules, and data processing. A limit of our study is that we smooth oscillations in time with a 64~s boxcar average, which effectively blocks us from detecting periods smaller than $\sim 120$~s. This might explain the larger value for the median period that we found compared to those in \citet{jafarzadeh17_2} and \citet{2021ApJ...921...30S}.

Observations of fibrils in H$\alpha$ and Type~II spicules show diverse velocity amplitudes, with median values ranging from 5~\kms\ to 10~\kms. The \CaIIK\ fibrils in this study, as well as those observed by \citet{jafarzadeh17_2}, have a POS velocity amplitude in the range of 1--2~\kms. This difference could be explained by the differences in formation height together with decreasing mass density with height: for constant wave flux (and phase speed), lower density gas implies larger velocity amplitude. 
For fibrils that appear both in \CaIIK\ and \Halpha, the segment of the fibril that appears bright in \CaIIK\ is likely located lower in the atmosphere than the segment only visible in H$\alpha$. \citet{Kianfar20} reported that \CaIIK\ shows fibrils located around $\log \tau_{500} \approx -4.3$, while the H$\alpha$ line core forms at larger heights ($\log \tau_{500} \approx -5.7$ in the FALC model atmosphere). Type~II spicules can protrude far above the bulk chromosphere and are expected to form at even lower densities. 


The presence of oscillations in both the LOS and POS direction, with comparable periods for 373 out of 468 fibrils in our sample, strongly suggests that both oscillation directions are excited by the same process. Theoretical calculations show that curved flux tubes have nearly the same wave speeds in the vertical and horizontal directions \citep{2009SSRv..149..299V}. 

The presence of fibrils with identical periods and zero phase difference in our sample is consistent with convective buffeting \citep{1981A&A....98..155S} with randomly oriented kicks as a driver because there is only a very weak correlation between the LOS and POS velocity amplitudes (Fig.~\ref{fig:A_P}). 
Simulations of photospheric magnetoconvection at high resolution
\citep[e.g.,][]{2014ApJ...789..132R}
show however substantially more complex motions of photospheric magnetic elements, which could explain the whole range of phase differences that we observe.

We find similar velocity amplitudes for the POS and the LOS oscillations. \citet{2021ApJ...921...30S} found smaller LOS amplitudes than POS amplitudes in \Halpha\ spicules. They defined the LOS velocity as the first moment of the line profile, which can yield different values than the actual LOS gas velocity
\citep[see Figs.~5 and~6 in][]{2022arXiv220813749D}.
Measurement of fibril oscillations in H$\alpha$ using the Doppler shift of the line profile minimum as a LOS velocity measurement \citep{2012ApJ...749..136L} could provide a more reliable measurement of the relative sizes of the LOS and POS velocity amplitudes.

The existence of fibrils with a phase difference of $\pi/2$ between the oscillations in POS and LOS directions indicates the presence of fully circularly polarized waves, in the context of flux tubes known as helical kink waves \citep{Zaqar2008}. This leads to circular motion of the tube axis in the plane perpendicular to the tube. Such motions have been detected before in both off-limb spicules
\citep[e.g.,][]{1982SvAL....8..341G} and on-disk spicules \citep{2021ApJ...921...30S}. Similar motions have also been detected in the chromosphere above photospheric magnetic elements using feature tracking \citep{2017ApJ...840...19S}. 


Finally, we note that estimates of the wave flux carried by transverse waves in chromospheric fibrils based on POS motions only are an underestimate because they do not include the significant LOS motions. As we show here, imaging spectroscopy together with non-LTE inversions is a powerful tool to study chromospheric oscillations in all three dimensions.

\begin{acknowledgements}
SK and JL were supported through the CHROMATIC project (2016.0019) funded by the Knut and Alice Wallenberg foundation. SEP was supported by the Knut and Alice Wallenberg Foundation and also acknowledges the funding received from the European Research Council (ERC) under the European Union’s Horizon 2020 research and innovation program (ERC Advanced Grant agreement No.742265) and from the Agencia Estatal de Investigación del Ministerio de Ciencia, Innovación y Universidades (MCIU/AEI) under grant ``Polarimetric Inference of Magnetic Fields'' and the European Regional Development Fund (ERDF) with reference PID2022-136563NB-I00/10.13039/501100011033. SD has received funding from Swedish Research Council (2021-05613) and Swedish National Space Agency (2021-00116). The Swedish 1-m Solar Telescope is operated on the island of La Palma by the Institute for Solar Physics of Stockholm University in the Spanish Observatorio del Roque de los Muchachos of the Instituto de Astrof{\'i}sica de Canarias. The Institute for Solar Physics is supported by a grant for research infrastructures of national importance from the Swedish Research Council (registration number 2021-00169). This project has received funding from the European Research Council (ERC) under the European Union's Horizon 2020 research and innovation program (SUNMAG, grant agreement 759548). The NSO is operated by the Association of Universities for Research in Astronomy, Inc., under cooperative agreement with the National Science Foundation.
\end{acknowledgements}

\bibliographystyle{aa} 
\bibliography{paperII}
\end{document}